\documentclass[11pt]{article}
\usepackage{cite}
\usepackage{mdframed}
\usepackage{epsfig,subfigure}
\usepackage{amssymb}
\usepackage[fleqn]{amsmath}
\usepackage{color}
\usepackage{arydshln}
\def\QED{\mbox{\rule[0pt]{1.5ex}{1.5ex}}}

\usepackage{graphicx}
\usepackage{color}
\usepackage[dvipsnames]{xcolor}

\definecolor{armygreen}{rgb}{0.29, 0.33, 0.13}

\usepackage{amssymb}
\usepackage{amsmath,amsfonts,amssymb}
\usepackage{verbatim}
\usepackage{stfloats}
\usepackage[bookmarks=false]{}
\usepackage{algorithm}
\usepackage{algcompatible}

\newtheorem{theorem}{Theorem}

\newtheorem{lemma}{Lemma}

\newtheorem{remark}{Remark}

\newtheorem{example}{Example}

\newcommand\blfootnote[1]{%
  \begingroup
  \renewcommand\thefootnote{}\footnote{#1}%
  \addtocounter{footnote}{-1}%
  \endgroup
}

\begin{document}
\date{}

\title{
Secure Groupcast with Shared Keys
}
\author{\normalsize Hua Sun \\
}

\maketitle

\blfootnote{
Hua Sun (email: hua.sun@unt.edu) is with the Department of Electrical Engineering at the University of North Texas. }

\maketitle

\begin{abstract}
We consider a transmitter and $K$ receivers, each of which shares a key variable with the transmitter. Through a noiseless broadcast channel, the transmitter wishes to send a common message $W$ securely to $N$ out of the $K$ receivers while the remaining $K-N$ receivers learn no information about $W$. We are interested in the maximum message rate, i.e., the maximum number of bits of $W$ that can be securely groupcast to the legitimate receivers per key block and the minimum broadcast bandwidth, i.e., the minimum number of bits of the broadcast information required to securely groupcast the message bits.

We focus on the setting of combinatorial keys, where every subset of the $K$ receivers share an independent key of arbitrary size. Under this combinatorial key setting, the maximum message rate is characterized for the following scenarios - 1) $N=1$ or $N=K-1$, i.e., secure unicast to 1 receiver with $K-1$ eavesdroppers or secure groupcast to $K-1$ receivers with $1$ eavesdropper, 2) $N=2, K=4$, i.e., secure groupcast to $2$ out of 4 receivers, and 3) the symmetric setting where the key size for any subset of the same cardinality is equal for any $N,K$. Further, for the latter two cases, the minimum broadcast bandwidth for the maximum message rate is characterized.
\end{abstract}

\newpage

\allowdisplaybreaks
\section{Introduction}
\begin{figure}[h]
\begin{center}
\includegraphics[width=5.5 in]{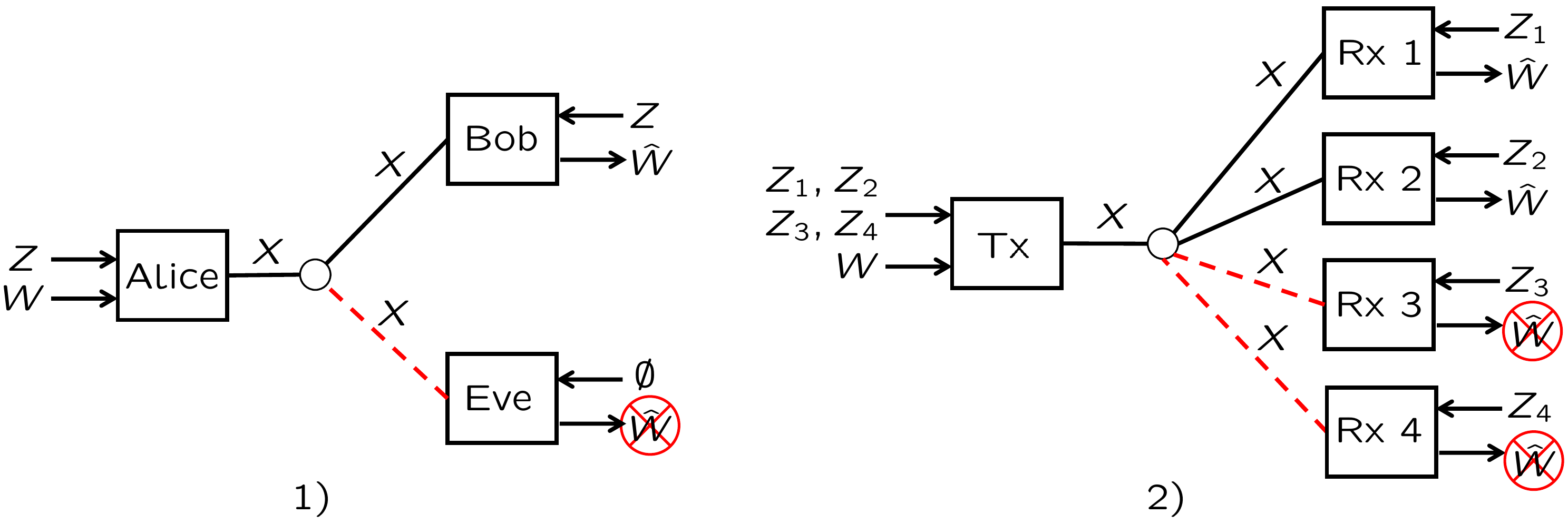}
\caption{\small 1) The one-time pad system. 2) The secure groupcast problem (to $2$ out of $4$ receivers).}
\label{fig:model}
\end{center}
\end{figure}

The first theoretical analysis of cryptography and secrecy system was carried out by Shannon in the groundbreaking 1949 work \cite{Shannon1949}, where the mathematical framework of information theoretic security was introduced to establish the fundamental limits of secure point-to-point communication. Shannon studied the one-time pad system (see Fig.~\ref{fig:model}.1), where Alice shares a key $Z$ with Bob and wishes to send an independent message $W$ to Bob such that even if the transmit signal $X$ is fully eavesdropped by Eve, Eve cannot learn anything about $W$ as long as Eve has no knowledge of the key $Z$. The simple one-time pad scheme $X = W + Z$, where `$+$' represents bit-wise binary addition is proved information theoretically secure and communication-wise optimal in the following sense.

\begin{itemize}
\item To send one bit of the message $W$ securely, one bit of the key $Z$ must be shared. That is, the maximum {\em message rate} is 1 bit per key bit.
\item To send one bit of the message $W$ securely, one bit of the transmit signal $X$ must be broadcast (seen by everyone). That is, the minimum {\em broadcast bandwidth} is 1 bit per message bit.
\end{itemize}

\vspace{-0.1in}
\vspace{-0.15in}

In this work, motivated by the need of secure {\em group} (beyond point-to-point) communication under {\em complex} adversarial scenario (beyond a single eavesdropper knowing nothing about the key), we consider the following secure groupcast communication scenario. A transmitter shares a key variable $Z_k, k \in \{1,2,\cdots, K\}$ with Receiver $k$ and $Z_k$ may be arbitrarily correlated (see Fig.~\ref{fig:model}.2). Aided by the shared keys, the transmitter wishes to send a common message $W$ securely to $N$ out of the $K$ receivers through broadcasting the signal $X$ to all receivers, in a way that any one of the remaining $K-N$ receivers learns no information about $W$ in the information theoretic sense.

This secure groupcast problem naturally generalizes Shannon's one-time pad system, which is a special case of secure unicast $(N=1)$ over a $K=2$ receiver broadcast channel and the eavesdropping receiver knows nothing about the key of the legitimate receiver. Following the communication metrics considered by Shannon, we focus on the following two questions regarding the fundamental limits of secure groupcast.
\begin{itemize}
\item What is the maximum message rate, defined as the maximum number of bits of the message $W$ that can be securely groupcast per key block (a classic Shannon theoretic formulation where we may code over a long key block and the block size is allowed to approach infinity)? 
\item What is the minimum broadcast bandwidth, defined as the minimum number of bits of the broadcast information $X$ required to securely groupcast a message of certain rate?
\end{itemize}
Beyond being an elemental model for information theoretic security, the above shared key secure groupcast problem arises naturally in many applications, where we interpret the keys either as digital tokens or information from memory devices (e.g., in premiere streaming or game distribution), 
or more generally as side-information variables that could be sensed from the environment or obtained from prior communication (e.g., in wireless networking). In addition, the model can be easily extended from groupcasting a single message for a single group to multiple messages, each exclusively for an arbitrary group under various security constraints, i.e., the secure groupcast model is introduced to enable broadcasting to a selected set of qualified receivers while unqualified receivers obtain no useful information.

\subsection*{Combinatorial Key Setting}
\begin{figure}[h]
\begin{center}
\includegraphics[width=3.75 in]{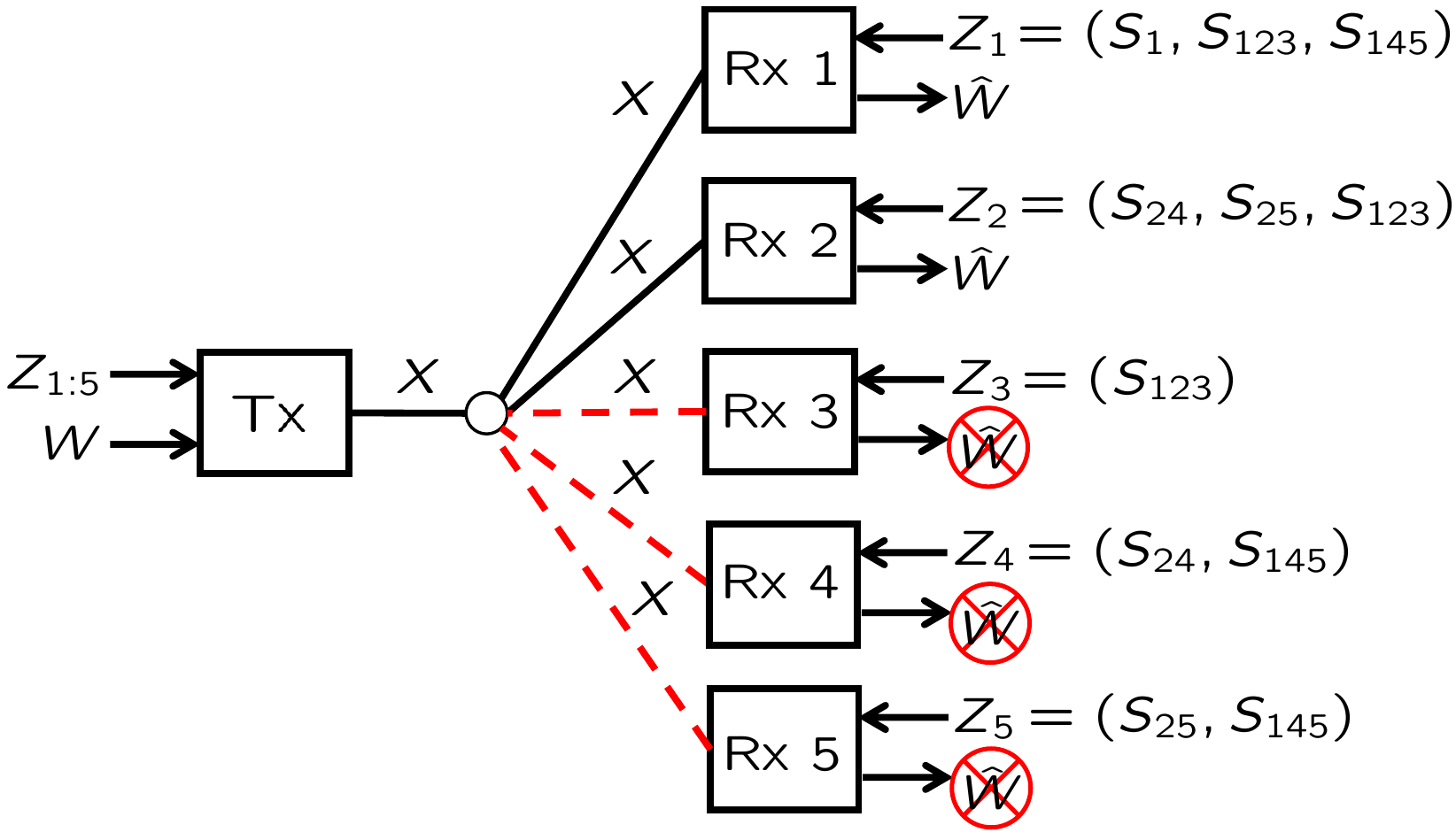}
\caption{\small A secure groupcast problem to $2$ out of $5$ receivers with combinatorial keys (i.e., the $S$ variables are independent). $Z_{1:5}$ denotes $(Z_1, Z_2, Z_3, Z_4, Z_5)$.}
\label{fig:52}
\end{center}
\end{figure}

As an initial step, we mainly focus on the combinatorial key setting, where every subset $\mathcal{U}$ of the $K$ receivers share an independent key $S_{\mathcal{U}}$ of arbitrary size. An example is shown in Fig.~\ref{fig:52}, where $S_1$ denotes the key that is known only to Receiver $1$ (and the transmitter), $S_{145}$ (abbreviation of $S_{\{1,4,5\}}$ for simplicity) is known to Receiver 1, Receiver 4 and Receiver 5 etc. Further, the $S$ variables with different subscripts are independent of each other.

\vspace{-0.15in}

The combinatorial key setting turns out to be technically challenging due to the necessity of highly structured coding of the message symbols and the key symbols (for which the setting in Fig.~\ref{fig:52} is a representative example even when the $S$ variables all have the same size), and the abundance of parameters (as the key size for different subsets may be different so that overall the order of parameters is exponential in $K$). The essence is to accommodate for and utilize the complex correlation among the keys so that legitimate receivers can decode the common message 
while eavesdropping receivers cannot obtain anything from the correlated keys (i.e., need to avoid leakage under multiple correlated views). 
Our results are summarized next.

\subsection*{Main Results and Techniques}
The main results of this work include the exact characterization of the maximum message rate and the minimum broadcast bandwidth for settings listed below.

\begin{itemize}
\item $N=1$, any $K$: This is the secure unicast setting, with only $1$ desired receiver. Both the maximum message rate and the minimum broadcast bandwidth are characterized. The achievable scheme is based on random linear coding over the key symbols. Refer to Theorem \ref{thm:unicast}.
\item $N=K-1$, any $K$: This can be viewed as the secure multicast setting, with only $1$ eavesdropper. The maximum message rate is characterized (and the minimum broadcast bandwidth when $K\leq 4$ or when the total secure key size is the same for all receivers). The achievable scheme is based on random linear coding over the message symbols. Refer to Theorem \ref{thm:multicast}.
\item $N=2, K=4$: Both the maximum message rate and the minimum broadcast bandwidth are characterized. The achievable scheme requires a delicate structured decomposition to basic components according to the key sizes. Refer to Theorem \ref{thm:42}.
\item The symmetric setting for any $N, K$, where the size of the key $S_{\mathcal{U}}$ only depends on $|\mathcal{U}|$ (i.e., the cardinality of the subset): Both the maximum message rate and the minimum broadcast bandwidth are characterized. The achievable scheme requires an intricate coding over keys from various subsets that handles correctness and security jointly. Refer to Theorem \ref{thm:sym}.
\item The converse bounds on the message rate for all results above are given by a simple conditional entropy term (refer to Theorem \ref{thm:msg}); the converse bounds on the broadcast bandwidth for all results above have an interesting unified form that can be interpreted through common information (refer to Theorem \ref{thm:bw}). 
\item The simple conditional entropy converse bound in Theorem \ref{thm:msg} is not tight in general. Specifically, a stronger bound is derived for the setting in Fig.~\ref{fig:52} when the $S$ variables have the same size and the maximum message rate is characterized with  a matching vector linear coding scheme (refer to Theorem \ref{thm:52}). 
\end{itemize}

We have also explored the generalization to the following scenarios.
\begin{itemize}
\item The rate region of secure groupcasting multiple messages. Specifically, we consider $2$ legitimate receivers with $3$ desired messages ($1$ for each individual receiver so that the other receiver learns nothing and a common message for both receivers) and all these $3$ messages must be kept fully secure to an eavesdropping receiver. Refer to Theorem \ref{thm:region}.

\item The discrete memoryless key setting. Interestingly, the scenarios where random linear codes suffice for the combinatorial key setting (i.e., $N=1$ and $N=K-1$) generalize fully to discrete memoryless keys by random binning. Refer to Theorem \ref{thm:dmc}.
\end{itemize}

We start with the problem statement and defer the discussion of related prior work in key agreement, 
latent capacity region, secure broadcasting, and secure index coding to Section \ref{sec:prior}.

\bigskip
{\it Notation: For positive integers $K_1, K_2, K_1 \leq K_2$, we use the notation $[K_1:K_2] = \{K_1, K_1+1,\cdots, K_2\}$. The notation $|\mathcal{U}|$ is used to denote the cardinality of a set $\mathcal{U}$ and the notation $|X|$ is used to denote the number of elements of a vector $X$.  
For a matrix ${\bf V}$, ${\bf V}(i,j)$ represents the element in the $i$-th row and $j$-th column. For two matrices ${\bf V}_1, {\bf V}_2$ (with the same number of columns), $[{\bf V}_1; {\bf V}_2]$ denotes the row stack of ${\bf V}_1, {\bf V}_2$. A binomial coefficient $\binom{K}{U}$ is defined as $0$ if $K < U$.
}

\section{Problem Statement}\label{sec:model}
Define $K$ discrete random variables $z_1, z_2,\cdots, z_K$ of finite cardinality, drawn from an arbitrary joint distribution $P_{z_1, z_2,\cdots, z_K}$. Following the convention, $Z_1, Z_2,\cdots, Z_K$ denote $L$ length extensions of $z_1, z_2,\cdots, z_K$, i.e., $Z_1, Z_2,\cdots, Z_K$ are sequences of length $L$, such that the sequence of tuples $[Z_1(l), Z_2(l), \cdots, Z_K(l)]_{l=1}^L$ is produced i.i.d. according to $P_{z_1, z_2,\cdots, z_K}$.

Consider a transmitter that knows the keys $Z_1, Z_2,\cdots, Z_K$, and $K$ receivers such that Receiver $k$ knows $Z_k, k \in [1:K]$. The transmitter wishes to send a common message $W$ securely to the first $N$ receivers, where $1\leq N \leq K-1$. The message $W$ consists of $L_W$ i.i.d. uniform symbols from a finite field\footnote{As usual for an information theoretic formulation, the actual size of the message is allowed to approach infinity. We allow the optimization of both parameters of the key block length $L$ and the field size $p$, to match the code dimensions and simplify the presentation of the coding scheme.} $\mathbb{F}_p$ for a prime power $p$, so $H(W) = L_W \log_2 p$ bits. We assume that the message $W$ is independent of the key variables $Z_1, Z_2,\cdots, Z_K$.
\begin{eqnarray}
I(W; Z_1, Z_2,\cdots, Z_K) = 0. \label{wzind}
\end{eqnarray}

The communication channel is a noiseless broadcast channel, i.e., the transmit signal $X$ is sent by the transmitter and seen by every receiver. To securely groupcast the message $W$, the transmit signal $X$ consists of $L_X$ symbols from $\mathbb{F}_p$.

From the transmit signal $X$ and the key $Z_k$, each legitimate receiver must be able to decode the message $W$, with probability of error $P_e$. The probability of error must approach zero as the key block length $L$ approaches infinity\footnote{If $P_e$ is required to be exactly zero, then the $o(L)$ term can be replaced with 0. The situation is similar if zero leakage instead of vanishing leakage is required in the security constraint (\ref{sec}).}. From Fano's inequality, we have
\begin{eqnarray}
\mbox{[Correctness]}~~~ H(W | X, Z_k) = o(L), ~\forall k \in [1:N] \label{corr}
\end{eqnarray}
where any function of $L$, say $f(L)$, is said to be $o(L)$ if $\lim_{L\rightarrow \infty} f(L)/L = 0$. From the transmit signal $X$ and the key $Z_k$, each eavesdropping receiver obtains a negligible amount of information about the message $W$.
\begin{eqnarray}
\mbox{[Security]}~~~ I(W; X, Z_k) = o(L),~\forall k \in [N+1:K]. \label{sec}
\end{eqnarray}

The groupcast {\em rate} characterizes how many bits of the message are securely groupcast per key block, and is defined as follows.
\begin{eqnarray}
R = \frac{H(W)}{L} = \frac{L_W\log_2 p}{L}. \label{rate}
\end{eqnarray}
A rate $R$ is said to be achievable if there exists a sequence of secure groupcast schemes (indexed by $L$), each of rate greater or equal to $R$, for which $P_e \rightarrow 0$ as $L \rightarrow \infty$ (i.e., the correctness constraint (\ref{corr}) and the security constraint (\ref{sec}) are satisfied). The supremum of achievable rates is called the capacity $C$.

The broadcast bandwidth $\beta(R)$ characterizes how many bits of the transmit signal are broadcast per key block to securely groupcast a message of rate $R$, and is defined as follows.
\begin{eqnarray}
\beta(R) = \frac{L_X\log_2 p}{L}. \label{bw}
\end{eqnarray}
The achievable broadcast bandwidth is defined similarly, i.e., broadcast bandwidth $\beta(R)$ is said to be achievable if there exists a sequence of secure groupcast schemes, each of rate greater than or equal to $R$ and each of broadcast bandwidth smaller than or equal to $\beta(R)$, for which  $P_e \rightarrow 0$ as $L \rightarrow \infty$. The infimum of achievable broadcast bandwidth is called the minimum broadcast bandwidth $\beta^*(R)$.

We will be mainly interested in the capacity, $C$ and the minimum broadcast bandwidth when the rate value is the capacity, $\beta^*(C)$.

\subsection{Combinatorial Keys}
The combinatorial key setting refers to a specific type of joint distribution of the keys and is defined as follows. Consider $2^K-1$ independent random variables $s_{\mathcal{U}}$, where $\mathcal{U}$ may be any non-empty subset of $[1:K]$. For example, when $K=3$, we have\footnote{For $s_{\mathcal{U}}$, we may simplify the subscript when the elements of $\mathcal{U}$ are easy to list, e.g., we may write $s_{\{1,2\}}$ as $s_{12}$.} $s_{1}, s_{2}, s_{3}, s_{12}, s_{13}, s_{23}, s_{123}$.
\begin{eqnarray}
H(s_1, s_2, \cdots, s_{\mathcal{U}}, \cdots, s_{1:K}) = H(s_1) + H(s_2) + \cdots + H(s_\mathcal{U}) + \cdots + H(s_{1:K}). \label{combi}
\end{eqnarray}
We assume that $s_{\mathcal{U}}$ consists of an integer number, say $L_{\mathcal{U}}$, of i.i.d. uniform symbols from $\mathbb{F}_p$.
\begin{eqnarray}
H(s_{\mathcal{U}}) = L_{\mathcal{U}} \log_2 p ~\mbox{bits}.
\end{eqnarray}
The variable $z_k$ is the collection of all $s_{\mathcal{U}}$ variables such that $k \in \mathcal{U}$.
\begin{eqnarray}
z_k = (s_{\mathcal{U}} : k \in \mathcal{U}).
\end{eqnarray}
For example, when $K=3$, $z_2 = (s_2, s_{12}, s_{23}, s_{123})$. The symmetric setting is defined as follows.
\begin{eqnarray}
\mbox{(symmetric setting)} ~~ H(s_{\mathcal{U}_1}) = H(s_{\mathcal{U}_2}), 
\forall \mathcal{U}_1, \mathcal{U}_2 ~\mbox{such that}~  |\mathcal{U}_1| = |\mathcal{U}_2|.
\end{eqnarray}

The extension of the above system model to include multiple groupcast messages is immediate and will be presented when we consider this generalization in Section \ref{sec:region}.

\section{Main Results}
In this section, we summarize our main results along with illustrative examples and observations.
\subsection{Converse on Rate $R$ and Broadcast Bandwidth $\beta(R)$}
We present a simple converse bound on the groupcast rate $R$ in the following theorem.

\begin{theorem}\label{thm:msg}
{\normalfont[Rate Converse]}
For any secure groupcast problem (to the first $N$ of $K$ receivers), 
\begin{eqnarray}
R \leq H(z_q | z_e), ~\forall q \in [1:N], \forall e \in [N+1:K]. \label{eq:msg}
\end{eqnarray}
\end{theorem}
The proof of Theorem \ref{thm:msg} is presented in Section \ref{sec:msg}. The conditional entropy bound (\ref{eq:msg}) is very intuitive, because $W$ must be decoded by any qualified Receiver $q \in [1:N]$ and cannot be learned by any eavesdropping Receiver $e \in [N+1:K]$. Surprisingly, this simple conditional entropy bound turns out to be tight for many settings of interest (see below). However, it is not sufficient in general (refer to Remark \ref{remark:nt} after Theorem \ref{thm:52}).

Next, we present an interesting converse bound on the broadcast bandwidth $\beta(R)$ in the following theorem. 

\begin{theorem}\label{thm:bw}
{\normalfont[Broadcast Bandwidth Converse]}
For any secure groupcast problem (to the first $N$ of $K$ receivers), consider any set of qualified receivers $\mathcal{Q} \triangleq \{q_1, \cdots, q_{|\mathcal{Q}|}\} \subset [1:N]$ and consider any random variable $u_e$ that is a function of the key of an eavesdropping Receiver $e \in [N+1:K]$, i.e., $H(u_e|z_e) = 0$.
\begin{eqnarray}
\beta(R) &\geq& |\mathcal{Q}| R - \left( \sum_{i=1}^{|\mathcal{Q}|} H(z_{q_i} | u_e) - H(z_{q_1}, z_{q_2}, \cdots, z_{q_{|\mathcal{Q}|}} | u_e) \right) \notag \\
&=&  |\mathcal{Q}| R - \sum_{i=1}^{|\mathcal{Q}|-1} I(z_{q_1}, \cdots, z_{q_i}; z_{q_{i+1}} | u_e). \label{eq:bw}
\end{eqnarray}
\end{theorem}

The proof of Theorem \ref{thm:bw} is presented in Section \ref{sec:bw}. The negative term on the RHS of (\ref{eq:bw}) captures the benefits of correlated keys in reducing the broadcast bandwidth. On one extreme when the keys are fully independent, this negative term is zero and we have to send the message $W$ to all $|\mathcal{Q}|$ qualified receivers one by one, so that the broadcast bandwidth is $|\mathcal{Q}|$ times of the rate, $R$ of the message, i.e., $\beta(R) \geq |\mathcal{Q}|R$. On the other extreme when the keys are identical and independent of the key at the eavesdropping receiver (i.e., $z_{q_1} = \cdots = z_{q_{\mathcal{Q}}}$), this negative term becomes $(|\mathcal{Q}|-1) H(z_{q_1})$. Now suppose $R = H(z_{q_1})$, then the broadcast bandwidth bound becomes $\beta(R) \geq R$ and it might suffice to simply send out the one-time pad signal $W + Z_{q_1}$. In general, between the two extremes, the saving is given by the difference between the sum of individual entropy of each key and the joint entropy of all keys, which can be interpreted as a form of common information. Interestingly, this common information type of term can also be written as the sum of a chain of mutual information terms.


Equipped with the above converse results, we are now ready to proceed to consider the combinatorial key setting, which is referred to as the combinatorial secure groupcast problem for short. Note that for combinatorial secure groupcast, all achievable schemes satisfy zero error and zero leakage (i.e., $o(L)$ is replaced with $0$ in (\ref{corr}), (\ref{sec})).

\subsection{Secure Unicast $N=1$ and Secure Multicast $N=K-1$ Settings}
When there is only $N= 1$ desired receiver, secure groupcast reduces to secure unicast, and this combinatorial secure unicast problem can be solved by random linear coding over the key symbols for the achievability side and the bounds given above for the converse side. This result is presented in the following theorem.

\begin{theorem}\label{thm:unicast}
{\normalfont[Secure Unicast]}
For the combinatorial secure unicast problem (to the first of $K$ receivers), the capacity and the minimum broadcast bandwidth for capacity achieving schemes are
\begin{eqnarray}
C &=& \min_{e\in [2:K]} H(z_1 | z_e) = \min_{e \in [2:K]} \sum_{\begin{subarray}{c}
\mathcal{U} \subset [1:K]: \\
1 \in \mathcal{U}, e \notin \mathcal{U}
\end{subarray}
} H(s_{\mathcal{U}}), \label{eq:unicast} \\
\beta^*(C) &=& C.
\end{eqnarray}
\end{theorem}

The proof of Theorem \ref{thm:unicast} is presented in Section \ref{sec:unicast}. The achievability is based on creating a key from what the legitimate Receiver 1 knows so that any one of the $K-1$ eavesdropping receivers cannot learn anything about the created key. This key can be created by random linear coding and after the key is created, one-time pad coding suffices to achieve the capacity and the minimum broadcast bandwidth. An example is presented below to explain this idea.

\begin{example}\label{ex:unicast}
Consider a combinatorial secure unicast instance with $K=4$ receivers, where the key configurations are given as follows. 
\begin{eqnarray}
z_1 = (s_{12}, s_{13}, s_{14}, s_{134}), ~z_2 = (s_{12}), ~z_3 = (s_{13}, s_{134}), ~z_4 = (s_{14}, s_{134}); \\
H(s_{12}) = 4 \log_2 p, ~H(s_{13}) = 2 \log_2 p, ~H(s_{14}) = \log_2 p, ~H(s_{134}) = 3 \log_2 p.
\end{eqnarray}
For example, $s_{12} \in \mathbb{F}_p^{4\times 1}$ contains $4$ symbols from $\mathbb{F}_p$.
From Theorem \ref{thm:unicast}, we have
\begin{eqnarray}
&& C = \min\Big(H(s_{13})+H(s_{14})+H(s_{134}), H(s_{12})+H(s_{14}), H(s_{12})+H(s_{13})\Big) = 5 \log_2 p , \\
&& \beta^*(C) = C = 5 \log_2 p.
\end{eqnarray}
The converse follows immediately from Theorem \ref{thm:msg} (taking the minimum converse bound over all eavesdropping receivers) and Theorem \ref{thm:bw} (taking $\mathcal{Q} = \{1\}$ so that $\beta(R) \geq R$). The achievable scheme is presented next. Consider $L=1$ block of the keys (then $Z_k = z_{k}$) and we wish to send $L_W = 5$ message symbols from $\mathbb{F}_p$ by broadcasting $L_X = 5$ symbols, i.e., $W, X$ are both $5\times 1$ vectors over $\mathbb{F}_p$. The combinatorial key variables are each precoded by a beamforming matrix to produce a mixed key, which is then added with the message $W$ to produce the transmit signal $X$.
\begin{eqnarray}
X = W + {\bf V}_{12} s_{12} + {\bf V}_{13} s_{13} + {\bf V}_{14} s_{14} + {\bf V}_{134} s_{134}
\end{eqnarray}
where the precoding matrices have $5$ rows each and the number of columns matches the dimension of the key variables, e.g., ${\bf V}_{12} \in \mathbb{F}_p^{5\times 4}$. The correctness constraint (\ref{corr}) is trivially satisfied. For eavesdropping Receiver $2$, after canceling the known key, he can recover
\begin{eqnarray}
W + {\bf V}_{13} s_{13} + {\bf V}_{14} s_{14} + {\bf V}_{134} s_{134}  = W + \Big[{\bf V}_{13} ~ {\bf V}_{14} ~{\bf V}_{134} \Big] 
\left[ \begin{array}{c}
s_{13} \\
s_{14} \\
s_{134}
\end{array}
\right]
\end{eqnarray}
so that in order to make sure nothing is revealed, we need 
\begin{eqnarray}
&& [{\bf V}_{13} ~{\bf V}_{14} ~{\bf V}_{134}]_{5\times 6} ~\mbox{to have full rank (Receiver 2)}.\\
\mbox{Similarly, we need} && [{\bf V}_{12} ~ {\bf V}_{14} ]_{5\times 5} ~\mbox{to have full rank (Receiver 3)}\\
and && [{\bf V}_{12} ~{\bf V}_{13}]_{5\times 6} ~\mbox{to have full rank (Receiver 4)}.
\end{eqnarray}
That is, we simply need the matrices to have full rank, which is easily satisfied by generic MDS matrices, e.g., Cauchy matrices over a properly large field. The details are deferred to the proof presented in Section \ref{sec:unicast}. Finally, the rate and broadcast bandwidth achieved match the converse.
\end{example}

We next consider the (somewhat) dual of secure unicast (a single legitimate receiver and any number of eavesdroppers) - secure multicast (a single eavesdropper and any number of legitimate receivers), whose capacity is solved by a similar random linear coding idea (but over the message symbols instead of over the key symbols). This result is presented in the following theorem.

\begin{theorem}\label{thm:multicast}
{\normalfont[Secure Multicast]}
For the combinatorial secure multicast problem (to the first $K-1$ of $K$ receivers), the capacity is 
\begin{eqnarray}
C = \min_{q\in[1:K-1]} H(z_q|z_K) = \min_{q\in[1:K-1]} \sum_{\mathcal{U} \subset [1:K-1]: q\in\mathcal{U}}
H(s_{\mathcal{U}}).
\end{eqnarray}
Further, the minimum broadcast bandwidth for capacity achieving schemes is characterized in the following two cases.
\begin{eqnarray}
&&1. ~H(z_{1}|z_K) = \cdots = H(z_{K-1}|z_K): ~\beta^*(C) = \sum_{\mathcal{U}\subset[1:K-1]} H(s_{\mathcal{U}}). \\
&&2. ~K=4 ~(\mbox{assume $H(z_1|z_4) \leq \min\Big(H(z_2|z_4), H(z_3|z_4)\Big), H(s_{12})\leq H(s_{13})$ with no loss}): \notag\\
&&\beta^*(C) =H(s_{123})+ \max\Big(2H(s_{1})+H(s_{12})+2H(s_{13}), 3H(s_1)+2H(s_{12})+2H(s_{13})-H(s_{23})\Big). \notag\\
\label{eq:multicast4}
\end{eqnarray}
\end{theorem}

The proof of Theorem \ref{thm:multicast} is presented in Section \ref{sec:multicast}. To achieve the capacity, we simply generate random linear combinations of the message symbols and mix them with each of the combinatorial keys. Each legitimate receiver can decode the message after collecting a sufficient number of coded message symbols. This idea is explained in the following example.

\begin{example}\label{ex:multicast}
Consider a combinatorial secure multicast instance with $K=4$ receivers, where the key configurations are given as follows. 
\begin{eqnarray}
z_1 = (s_{1}, s_{13}), ~z_2 = (s_{23}), ~z_3 = (s_{13}, s_{23}), ~z_4=(); \\
H(s_{1}) = \log_2 p, ~H(s_{13}) = 2 \log_2 p, ~H(s_{23}) = 3 \log_2 p.
\end{eqnarray}
From Theorem \ref{thm:multicast}, we have
\begin{eqnarray}
&&C = \min\Big(H(s_1) + H(s_{13}), H(s_{23}), H(s_{13}) + H(s_{23})\Big) = 3\log_2 p,\\
&& \beta^*(C) = \max\Big(2H(s_{1})+2H(s_{13}), 3H(s_1)+2H(s_{13})-H(s_{23})\Big) = 6 \log_2 p.
\end{eqnarray}
The rate converse is simply given by the minimum entropy of the legitimate key variables (and follows from Theorem \ref{thm:msg}). The broadcast bandwidth converse is given by Theorem 2, where $\mathcal{Q} = \{1,2\}$ and $u_e = ()$ so that $\beta(C) \geq 2C - I(z_1;z_2) = 2C$. The achievable scheme is presented next. Consider $L=1$ and $W \in \mathbb{F}_p^{3 \times 1}$. The transmit signal $X\in\mathbb{F}_p^{6\times 1}$ is produced as follows.
\begin{eqnarray}
X = \left[\begin{array}{c}
{\bf V}_1 W + s_1\\
{\bf V}_{13} W + s_{13}\\
{\bf V}_{23} W + s_{23}
\end{array}
\right]
\end{eqnarray}
where the dimensions of the precoding matrices are specified as ${\bf V}_1 \in \mathbb{F}_p^{1\times 3}, {\bf V}_{13} \in \mathbb{F}_p^{2\times 3}, {\bf V}_{23} \in \mathbb{F}_p^{3\times 3}$. For the secure multicast problem, security is trivial as the keys known by the eavesdropping receiver are never used and correctness requires that 
\begin{eqnarray}
&& [{\bf V}_1; {\bf V}_{13}] ~\mbox{has full rank (Receiver $1$)}, ~[{\bf V}_{23}] ~\mbox{has full rank (Receiver $2$)}, \\
&& \mbox{and}~[{\bf V}_{13}; {\bf V}_{23}] ~\mbox{has full rank (Receiver $3$)}.
\end{eqnarray}
The above constraints can be satisfied by generic (e.g., Cauchy or any MDS) matrices and details are deferred to the full proof presented in Section \ref{sec:multicast}. 
\end{example}

\begin{remark}\label{remark:multicastcom}
While the capacity of secure multicast is solved simply by random linear codes, the minimum broadcast bandwidth for capacity achieving schemes is generally an open problem (e.g., when $K\geq 5$). When $K\leq 4$, we need to analyze carefully which combinatorial key has redundancy and quantify the amount so as to use only the minimum required (see Section \ref{sec:mulbw}). 
\end{remark}

Interestingly, the above random coding idea for both combinatorial secure unicast and secure multicast generalizes to the discrete memoryless key setting, where corresponding results are obtained using standard existing random binning arguments (see Theorem \ref{thm:dmc}). 

\subsection{Secure Groupcast to $N=2$ of $K=4$ Receivers}
The capacity and minimum broadcast bandwidth for the combinatorial secure groupcast problem to $2$ out of $4$ receivers are characterized in the following theorem.
\begin{theorem}\label{thm:42}
{\normalfont[$N=2, K=4$]} 
For the combinatorial secure groupcast problem to the first $N=2$ of $K=4$ receivers, the capacity is
\begin{eqnarray}
C = \min_{q\in\{1,2\}, e\in\{3,4\}} H(z_q | z_e) = H(s_{12}) + \min \Big( H(s_1)+H(s_{14}) + H(s_{124}), H(s_1)+H(s_{13}) + H(s_{123}), \notag\\
H(s_2)+H(s_{24}) + H(s_{124}), H(s_2)+H(s_{23}) + H(s_{123}) \Big) \label{eq:42msg}
\end{eqnarray}
and the minimum broadcast bandwidth for capacity achieving schemes is 
\begin{eqnarray}
\beta^*(C) = 2C - H(s_{12}) - \min \Big( H(s_{123}), H(s_{124}) \Big). \label{eq:42bw}
\end{eqnarray}
\end{theorem}

The proof of Theorem \ref{thm:42} is presented in Section \ref{sec:42}. The complexity mainly lies in the abundance of the parameters so that we need to decompose the problem instance into multiple basic components (also how to identify basic components) and depending on the key configurations, there are many case studies. To this end, we need a decomposition result of two achievable schemes with two independent sets of keys, stated in the following lemma. 
\begin{lemma}\label{lemma:decompose}
Consider two sets of independent keys $(Z_1^{[1]}, \cdots, Z_K^{[1]})$ and $(Z_1^{[2]}, \cdots, Z_K^{[2]})$ such that $L_W^{[1]}$ and $L_W^{[2]}$ symbols of the messages $W^{[1]}, W^{[2]}$ can be securely groupcast with $L_X^{[1]}$ and $L_X^{[2]}$ symbols of the transmit signals $X^{[1]}, X^{[2]}$, respectively. Then we can concatenate the two schemes to one such that the keys are $Z_k = (Z_k^{[1]}, Z_k^{[2]}), k \in [1:K]$, $L_W = L_W^{[1]} + L_W^{[2]}$ symbols of $W = (W^{[1]}, W^{[2]})$ are securely groupcast with $L_X = L_X^{[1]} + L_X^{[2]}$ symbols of $X = (X^{[1]}, X^{[2]})$.
\end{lemma}
{\it Proof:} The proof is almost immediate. As long as each component scheme is correct and secure, the concatenated scheme will be correct and secure as the keys are independent and the message and transmit signal symbols are also independent. Further, this concatenation generalizes trivially to any number of independent key sets.
\hfill\QED

We are now ready to give an example of the combinatorial secure groupcast problem to $2$ of $4$ receivers, to illustrate the main idea.
\begin{example}\label{ex:42}
Consider a combinatorial secure groupcast instance to $2$ of $4$ receivers, where the key configurations are given as follows. Remember that $z_k = (s_\mathcal{U}: k \in \mathcal{U}), k \in \{1,2,3,4\}$.
\begin{eqnarray}
(H(s_1), H(s_2), H(s_{13}), H(s_{14}), H(s_{23}), H(s_{24}), H(s_{123}), H(s_{124})) = (1,2,2,3,1,2,2,1).
\end{eqnarray}
From Theorem \ref{thm:42}, we have
\begin{eqnarray}
&&C = \min_{q\in\{1,2\}, e\in\{3,4\}} H(z_q | z_e) = \min(5, 5, 5, 5) = 5,\\
&& \beta^*(C) = 2C - 0 - \min(2,1) = 9.
\end{eqnarray}
The rate converse follows from the conditional entropy bound in Theorem \ref{thm:msg} and the broadcast bandwidth converse follows from Theorem \ref{thm:bw} by setting $\mathcal{Q} = \{1,2\}$ and $u_e = z_3$ or $z_4$. The achievable scheme is shown in the following figure, where we decompose the instance into 3 sub-networks. We operate over the binary field $\mathbb{F}_2$, i.e., $p=2$ and key block size is $L=1$. The total number of bits in the message and the transmit signal match the converse above ($5$ and $9$, respectively). All key bits are used, e.g., $H(s_{14}) = 3$ so that we have $3$ bits of $s_{14}$, and sub-network $1$ uses $1$ bit and sub-network $3$ uses $2$ bits (see Fig.~\ref{fig:ex42}). Correctness and security are easy to verify.

\begin{figure}[h]
\begin{center}
\includegraphics[width=6 in]{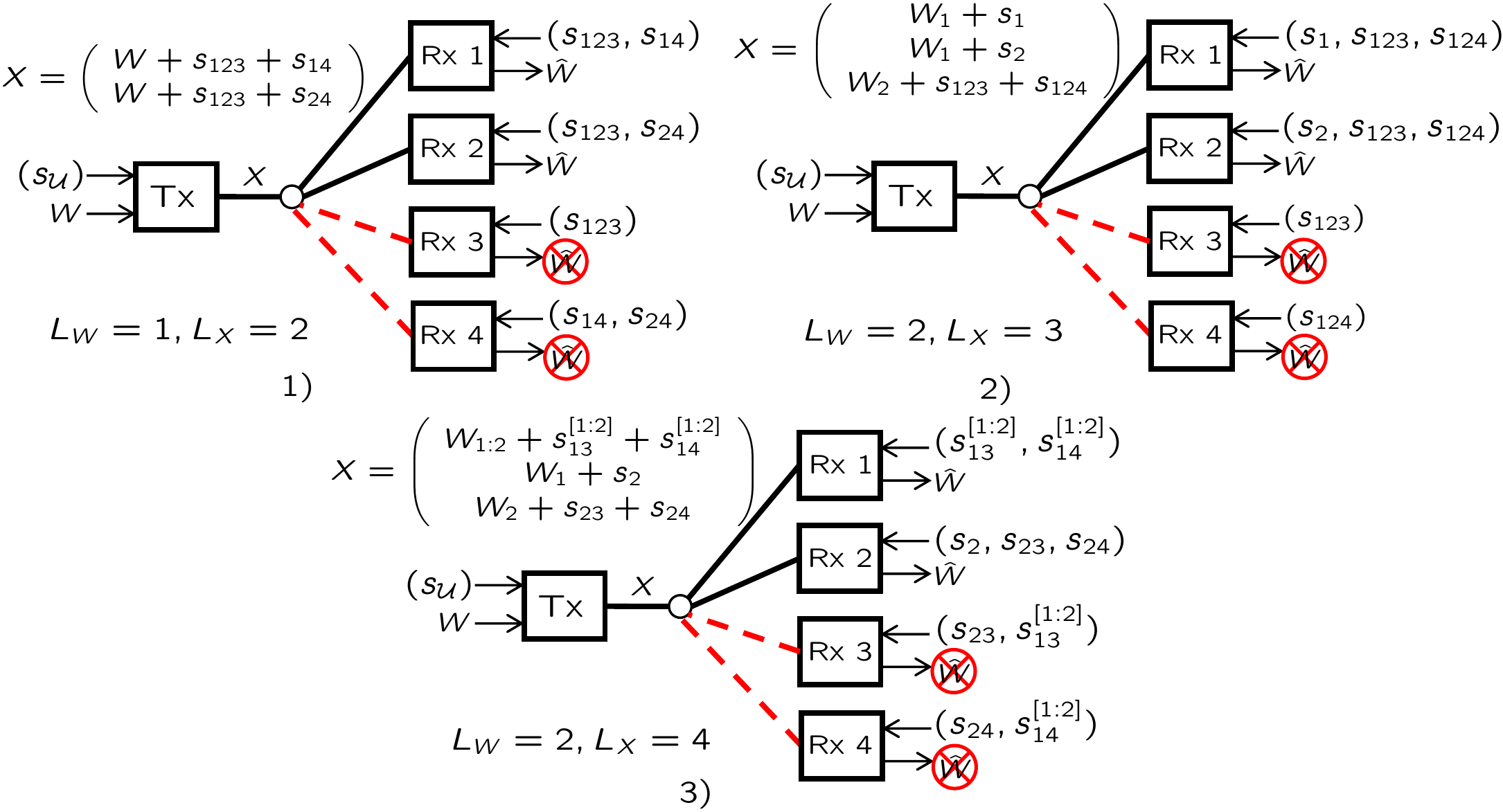}
\vspace{-0.1in}
\caption{\small The $3$ sub-networks, where the variables are independent for different sub-networks. When some variable has $2$ symbols, it is denoted as $W = (W_1; W_2) \triangleq W_{1:2}, s_{13} = (s_{13}^{[1]}; s_{13}^{[2]}) \triangleq s_{13}^{[1:2]}$ etc.}
\label{fig:ex42}
\end{center}
\end{figure}
\end{example}
\vspace{-0.45in}

\subsection{Secure Groupcast: Symmetric Setting}
We consider now the symmetric setting, where the key size only depends on the cardinality of the set of the receivers that have the same key. For any set $\mathcal{U} \subset [1:K]$ with cardinality $|\mathcal{U}| = u, u \in [1:K]$, we denote the key size as $H(s_{\mathcal{U}}) = L^{[u]} \log_2 p$. The capacity and minimum broadcast bandwidth for the symmetric setting are characterized in the following theorem.

\begin{theorem}\label{thm:sym}
{\normalfont[Symmetric Setting]}
For the symmetric combinatorial secure groupcast problem (to the first $N$ of $K$ receivers), the capacity and the minimum broadcast bandwidth for capacity achieving schemes are
\begin{eqnarray}
C = \sum_{u=1}^K \binom{K-2}{u-1}L^{[u]}\log_2 p, ~~\beta^*(C) = \sum_{u=1}^K \left( \binom{K-1}{u} - \binom{K-N-1}{u} \right) L^{[u]} \log_2 p.
\end{eqnarray}
\end{theorem}

We refer to a key that is known to $u$ receivers as a $u$-key. From the capacity and broadcast bandwidth formula, we see that it suffices to consider  $u$-keys separately for distinct $u$ values, i.e., joint coding across different $u$-keys is not necessary. This is generally not true (e.g., see Fig.~\ref{fig:ex42}.1) and greatly simplifies the problem. After we notice this simplification (we may limit to only $u$-keys of one $u$ value), the problem still requires an intricate decomposition of the keys, depending on how many qualified receivers and how many eavesdropping receivers know the key. An example is presented below to illustrate the main idea and the full proof is presented in Section \ref{sec:sym}.

\begin{example}\label{ex:sym}
Consider a symmetric combinatorial secure groupcast instance to $N=3$ of $K=6$ receivers, where we only have $3$-keys, i.e., $\binom{6}{3}$ keys of the same length $L^{[3]} = 1$. 

From Theorem \ref{thm:sym}, we have
\begin{eqnarray}
C = \binom{4}{2} \log_2 p = 6\log_2 p, ~~\beta^*(C) = \binom{5}{3} \log_2 p = 10 \log_2 p.
\end{eqnarray}
The rate converse follows from the conditional entropy bound in Theorem \ref{thm:msg}, where we may pick any qualified receiver and any eavesdropping receiver such that the qualified receiver knows $\binom{4}{2}$ keys that are not known to the eavesdropping receiver. The broadcast bandwidth converse follows from Theorem \ref{thm:bw} by setting $\mathcal{Q} = [1:3]$ and $u_e = z_6$ so that $\beta(C) \geq  H(z_1, z_2, z_3 | z_6) = \binom{5}{3} \log_2 p$.

The achievability is designed based on dividing the $3$-keys into $3$ groups. 
\begin{enumerate}
\item The first group involves the key that is known only to $3$ qualified receivers, i.e., $s_{123}$. As $s_{123}$ is not known to the eavesdropping receivers, we simply send $1$ message symbol with $1$ symbol of one-time pad transmit signal, i.e., we have achieved $R^{1} = \beta^{[1]}(R^{1}) = \log_2 p$. 
\item The second group involves the keys that are known to $2$ qualified receivers and $1$ eavesdropping receiver. We need to further divide these keys depending on the set of $2$ qualified receivers. Suppose the set of qualified receivers is $\{1,2\}$, i.e., we are considering the keys $(s_{124}, s_{125}, s_{126})$ that are common to qualified Receiver $1$ and qualified Receiver $2$. Further, any eavesdropping receiver only knows $1$ key from $(s_{124}, s_{125}, s_{126})$. In other words, we have the secure unicast situation (note that here Receiver $1$ and Receiver $2$ both require the same message and hold the same key) where the desired receiver has $4$ equal-size combinatorial key variables while the eavesdropping receivers have $1$ combinatorial key each. Therefore, combining generic linear coding ideas for key symbols in Theorem \ref{thm:unicast} and for message symbols in Theorem \ref{thm:multicast}, we can send $3-1=2$ symbols securely to Receiver $1$ and Receiver $2$ by transmitting
\begin{eqnarray}
X^{2,12} = {\bf V}^w_{12} W^{2} + {\bf V}^s_{124} s_{124} + {\bf V}^s_{125} s_{125} + {\bf V}^s_{126} s_{126}
\end{eqnarray} 
where ${\bf V}^s_{124}, {\bf V}^s_{125}, {\bf V}^s_{126} \in \mathbb{F}_p^{2\times 1}$ and ${\bf V}^{w}_{12} \in \mathbb{F}_p^{2\times 4}$ (the reason of setting this size will be clear soon).
We repeat the same coding procedure for the other $\binom{3}{2} - 1 = 2$ sets of keys, i.e., $(s_{134}, s_{135}, s_{136})$ (common keys to qualified Receiver $1$ and qualified Receiver $3$) and $(s_{234}, s_{235}, s_{236})$ (common to Receiver $2$ and Receiver $3$).
\begin{eqnarray}
X^{2,13} = {\bf V}^w_{13} W^{2} + {\bf V}^s_{134} s_{134} + {\bf V}^s_{135} s_{135} + {\bf V}^s_{136} s_{136} \\
X^{2,23} = {\bf V}^w_{23} W^{2} + {\bf V}^s_{234} s_{234} + {\bf V}^s_{235} s_{235} + {\bf V}^s_{236} s_{236} 
\end{eqnarray} 
where ${\bf V}^s \in \mathbb{F}_p^{2\times 1}, {\bf V}^w \in \mathbb{F}_p^{2\times 4}$. From the transmit signal $X^2 = (X^{2,12}, X^{2,13}, X^{2,23})$, each qualified receiver can obtain $4$ generic desired message combinations (so the size of ${\bf V}^w$ is chosen to match this total number of combinations), from which $4$ symbols of $W^2$ can be recovered as long as the ${\bf V}^w$ matrices are chosen in a generic manner. For example, qualified Receiver $1$ can obtain $({\bf V}^w_{12} W^{2}, {\bf V}^w_{13} W^{2})$. Security is guaranteed as long as the ${\bf V}^s$ matrices are chosen generically so that eavesdropping receivers see a sufficiently number of generic key combinations. 
To sum up, the overall rate and broadcast bandwidth achieved for all keys in the second group are
\begin{eqnarray}
R^2 = \binom{3-1}{2-1}2\log_2 p = 4\log_2 p, ~~\beta^2(R^2) = \binom{3}{2} 2 \log_2 p = 6\log_2p.
\end{eqnarray}
\item The third group involves the keys that are known to $1$ qualified receiver and $2$ eavesdropping receivers. We need to further divide these keys depending on the identity of the qualified receiver. Suppose the qualified receiver is Receiver $1$, i.e., we are considering the keys $(s_{145}, s_{146}, s_{156})$ such that any eavesdropping receiver knows $2$ of these $3$ keys (e.g., eavesdropping Receiver $4$ knows $s_{145}, s_{146}$). From the result of secure unicast (refer to Theorem \ref{thm:unicast}), we can achieve $R = \beta(R) = (2-1)\log_2 p = \log_2 p$ for these $3$ keys. Repeat the same procedure for $(s_{245}, s_{246}, s_{256})$ and $(s_{345}, s_{346}, s_{356})$. The overall rate and broadcast bandwidth achieved for all keys in the third group are
\begin{eqnarray}
R^3 = \log_2 p, ~~\beta^3(R^3) = 3 \log_2 p.
\end{eqnarray}
\end{enumerate}
Finally, we combine the performance of all $3$ groups (using Lemma \ref{lemma:decompose} for independent keys), so the total rate and broadcast bandwidth achieved are
\begin{eqnarray}
R = R^1 + R^2 + R^3 = 6 \log_2 p, ~~\beta(R) = \beta^1(R^1) +  \beta^2(R^2) +  \beta^3(R^3) = 10\log_2 p
\end{eqnarray}
which match the converse.
\end{example}

\subsection{A Secure Groupcast Instance with $N=2, K=5$}
\begin{figure}[h]
\begin{center}
\includegraphics[width=5 in]{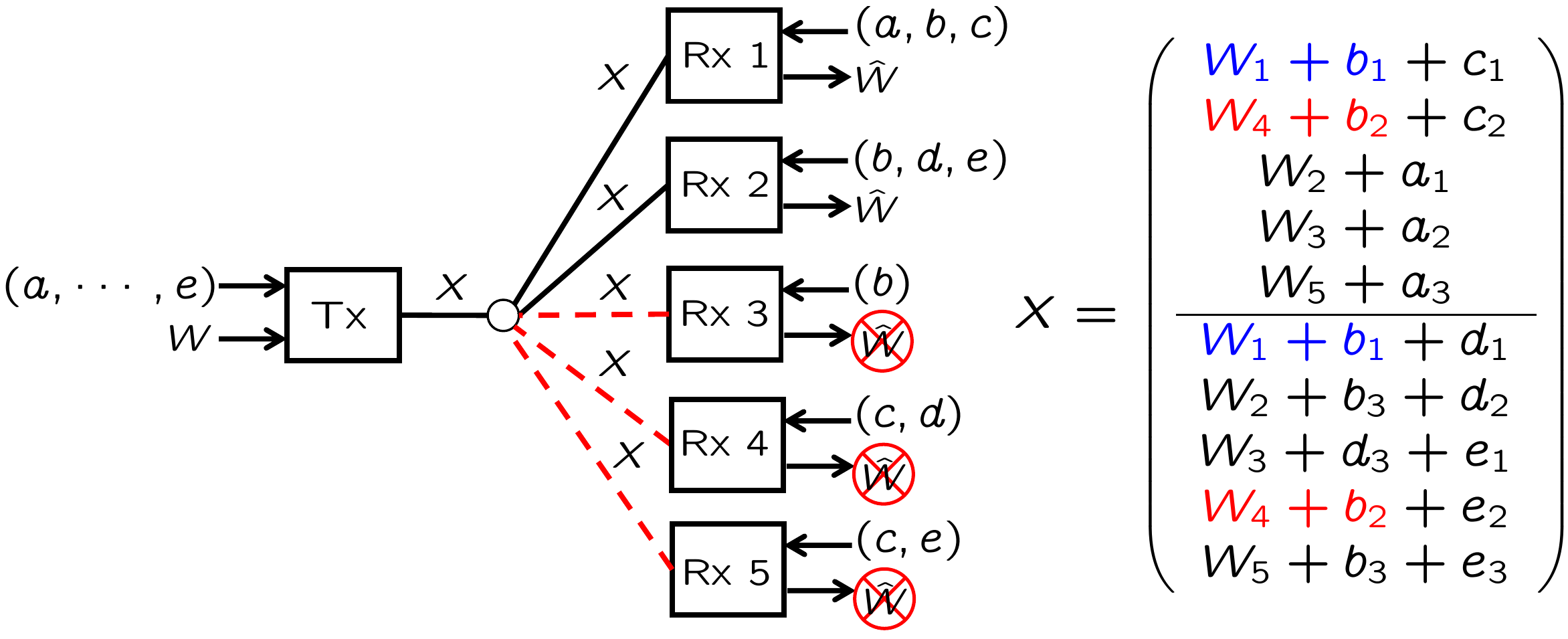}
\caption{\small A combinatorial secure groupcast instance to $2$ of $5$ receivers. The keys $a,b,c,d,e$ are independent and each key has $1$ bit per block. The message has $5$ bits, $W = (W_1, \cdots, W_5)$, sent over $3$ key blocks so that each key has $3$ bits, e.g., $a = (a_1,a_2,a_3)$. The broadcast signal $X$ has $10$ bits.
}
\label{fig:52re}
\end{center}
\end{figure}

For all capacity results presented above, the conditional entropy converse bound in Theorem \ref{thm:msg} turns out to be tight. We wonder if the bound is always tight. Along this line, we find that the answer is negative. We identify a simplest setting of combinatorial secure groupcast instance to $N=2$ of $K=5$ receivers (note that all settings with smaller $N,K$ values are settled by the converse bound in Theorem \ref{thm:msg}) where a strictly stronger converse is required. The setting turns out to be that in Fig.~\ref{fig:52} and is redrawn here with simplified notations (refer to Fig.~\ref{fig:52re}). We have characterized its capacity and minimum broadcast bandwidth, and this result is presented in the following theorem.

\vspace{-0.35in}

\begin{theorem}\label{thm:52}
For the combinatorial secure groupcast instance in Fig.~\ref{fig:52re}, the capacity and the minimum broadcast bandwidth for capacity achieving schemes are
\begin{eqnarray}
C = 5/3, ~~\beta^*(C) = 10/3. 
\end{eqnarray}
\end{theorem}

The achievable scheme is shown in Fig.~\ref{fig:52re}. Correctness is easy to verify, e.g., qualified Receiver $1$ knows $a, b, c$ such that from the first $5$ rows of the transmit signal $X$, he can obtain all message bits $(W_1, W_2, W_3, W_4, W_5)$. Security is more interesting. Eavesdropping Receiver $3$ learns nothing because even if $b$ is known, all message bits in $X$ are protected by $c, a, d, e$ (not known to Receiver $3$). Eavesdropping Receiver $4$ knows $c,d$ and after canceling $c,d$, he can obtain 
${\color{blue}W_1 +  b_1}, W_4 + b_2 , {\color{blue}W_1 + b_1}, W_2 + b_3$ that contains $b$.
However, nothing is leaked because the first term is the same as the third term (highlighted in blue). 
In fact, this is the key of the design. Therefore,
\begin{eqnarray}
I(W; X, c, d) &\overset{(\ref{wzind})}{=}& I(W; X | c, d) \\
&=& H(X | c,d) - H(X|W, c, d) = 9 - 9 = 0.
\end{eqnarray}
The situation for eavesdropping Receiver $5$ is similar, where the same noise of $b_2$ (noise alignment) appears in the same red signal ${\color{red}W_4+b_2}$ (signal alignment). Interestingly, similar alignment view has been proved useful recently in several other security and privacy primitives \cite{Li_Sun_CDS, Zhou_Sun_Fu, Sun_Jafar_PIR}.

We now discuss the rate converse. Here we give an intuitive argument for linear schemes, which guides the design of the achievable scheme, and defer the information theoretic proof to Section \ref{sec:52}, which is based on the sub-modularity property of entropy functions. Consider qualified Receiver $1$, who knows only $a, b, c$ and can decode $W$. Then $W$ must be fully recoverable from the key variables $a, b, c$ in $X$. As $a$ is only known to Receiver $1$, it can easily be used to transmit $L$ message bits. Then to achieve rate $R$, the message bits carried by $b,c$ must be $(R-1)L$ bits. As $b$ and $c$ are known to eavesdropping Receiver 3 and eavesdropping Receiver 4, respectively, the $(R-1)L$ message bits must be protected by both $b$ and $c$. Denote these $(R-1)L$ dimensions of $b$ as $B_1$. Now consider qualified Receiver $2$, who knows $b,d,e$ such that there must exist $RL$ dimensional space of $W$ that is covered by $b, d, e$. These $RL$ dimensions must be fully covered by $d,e$ as $b$ is known to eavesdropping Receiver $3$. As $d, e$ have dimension $L$ each, their overlap is $(2 - R)L$ and each of them separately covers $\frac{1}{2}(R - (2-R))L = (R-1)L$ dimensions. Therefore, eavesdropping Receiver $4$ can fully recover the $(R-1)L$ dimensions covered only by $d$ (and $b$) as $d$ is known. This $(R-1)L$ dimensional space of $b$ is denoted as $B_2$. Symmetrically, the $(R-1)L$ space of $b$ (mixed with $e$) after $e$ is known is denoted as $B_3$. Finally, we connect $B_1, B_2, B_3$. The desired message bits in $B_2, B_3$ are independent, so $B_1 \cap (B_2 \cap B_3) = \emptyset$. Otherwise, in $B_1$, we have the same $b$ space ($B_2 \cap B_3$) mixed with different desired message bits (security violated). Then
$L \geq \dim(B_1) + \dim(B_2 \cap B_3) \geq (R-1)L + (2(R-1) - 1)L = (3R - 4)L$, and $3R \leq 5$.
We note that the translation of this linear argument into an information theoretic proof with entropy terms is highly non-trivial.
The converse for the broadcast bandwidth is immediate, by setting $\mathcal{Q} = \{1,2\}$ and\footnote{Here we slightly abuse the notation. Note that in $u_e$, $e$ denotes the index of an eavesdropping receiver and not the combinatorial key $e$ in the secure groupcast instance.} $u_e = z_3 = b$ in Theorem \ref{thm:bw}: $\beta(C) \geq 2C - I(a, b, c; b, d, e | b) = 2C - I(a,c;d,e | b) = 2C = 10/3$.

\begin{remark}\label{remark:nt}
The conditional entropy converse bound in Theorem \ref{thm:msg} is $R \leq 2$ for the secure groupcast instance in Fig.~\ref{fig:52re}, which is strictly weaker than the capacity $5/3$. Thus the conditional entropy converse bound is not tight in general, for combinatorial secure groupcast (and  for secure groupcast).
\end{remark}

\section{Generalizations}
In this section, to show how insights generalize, we consider two extensions of the basic combinatorial secure groupcast model - the first one includes multiple messages and in the second one, keys are discrete memoryless sources.

\subsection{Secure Groupcasting Multiple Messages}\label{sec:region}
\begin{figure}[h]
\begin{center}
\includegraphics[width=4 in]{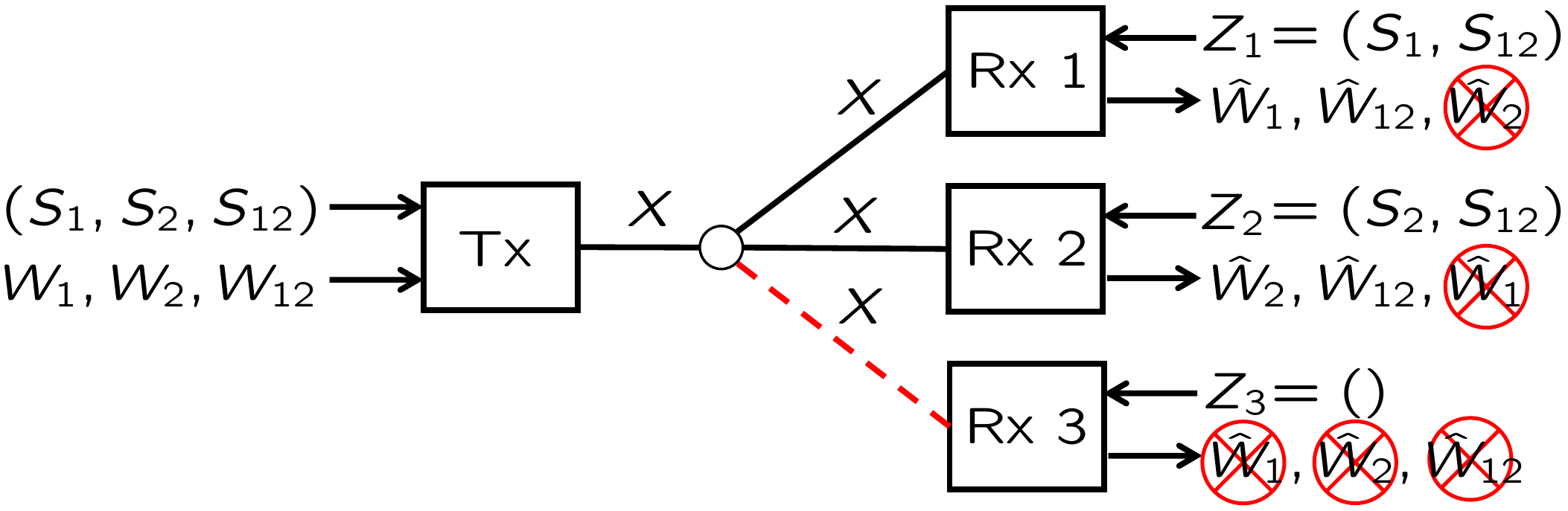}
\caption{\small A secure groupcast problem with $3$ messages and $3$ receivers.}
\label{fig:region}
\end{center}
\end{figure}

We consider an elementary $3$ receiver broadcast network with $3$ messages (see Fig.~\ref{fig:region}).
\vspace{-0.2in}

We first succinctly describe the model, which generalizes that in Section \ref{sec:model}. A transmitter wishes to deliver $3$ messages $W_1, W_2, W_{12}$ (of size $L_{W_1}, L_{W_2}, L_{W_{12}}$ i.i.d. uniform bits, respectively) through broadcasting a signal $X$ of size $L_X$ bits such that Receiver $1$ only learns $W_1, W_{12}$, Receiver 2 only learns $W_2, W_{12}$, and Receiver 3 learns nothing. Receiver $1$ is equipped with key $Z_1 = (S_1, S_{12})$, Receiver 2 is equipped with key $Z_2 = (S_2, S_{12})$ and Receiver 3's key is empty. The key variables $S_1, S_2, S_{12}$ are $L$ length extensions of uniform bits $s_1, s_2, s_{12}$ (of size $L_{1}, L_2, L_{12}$ bits, respectively).
\begin{eqnarray}
H(W_1, W_2, W_{12}, s_1, s_2, s_{12}) = H(W_1) + H(W_2) + H(W_{12}) + H(s_1) + H(s_2) + H(s_{12}). \label{eq:regioni}
\end{eqnarray}
The correctness and security constraints are as follows.
\begin{eqnarray}
\mbox{(Receiver 1)} && H(W_1, W_{12} | X, S_1, S_{12}) = o(L), ~I(W_2; X, S_1, S_{12}) = o(L) \label{eq:region1}\\
\mbox{(Receiver 2)} && H(W_2, W_{12} | X, S_2, S_{12}) = o(L), ~I(W_1; X, S_2, S_{12}) = o(L) \label{eq:region2}\\
\mbox{(Receiver 3)} && I(W_1, W_2, W_{12}; X ) = o(L). \label{eq:region3}
\end{eqnarray}
The rate of the messages and the broadcast bandwidth are defined as follows.
\begin{eqnarray}
R_1 = \frac{L_{W_1}}{L}, ~R_2 = \frac{L_{W_2}}{L}, ~R_{12} = \frac{L_{W_{12}}}{L}, ~\beta(R_1, R_2, R_{12}) = \frac{L_X}{L}.
\end{eqnarray}
The closure of the set of achievable rate tuples $(R_1, R_2, R_{12})$ is called the capacity region $\mathcal{C}$ and the minimum broadcast bandwidth for a rate tuple $(R_1, R_2, R_{12})$ is denoted as $\beta^*(R_1, R_2, R_{12})$.

We present the capacity region and the minimum broadcast bandwidth for the $3$ message secure groupcast problem in Fig.~\ref{fig:region} in the following theorem.

\begin{theorem}\label{thm:region}
{\normalfont[Rate Region]}
For the $3$ message combinatorial secure groupcast problem in Fig.~\ref{fig:region}, the capacity region and the minimum broadcast bandwidth are
\begin{eqnarray}
&& 0\leq R_1 + R_{12} \leq H(s_1) + H(s_{12}) \label{eq:regionm1}\\
&& 0\leq R_2 + R_{12} \leq H(s_2) + H(s_{12}) \label{eq:regionm2}\\
&& 0\leq R_1 \leq H(s_1) \label{eq:regionm3}\\
&& 0\leq R_2 \leq H(s_2) \label{eq:regionm4}\\
&& \beta^*(R_1, R_2, R_{12}) = R_1 + R_2 + \max(R_{12}, 2R_{12} - H(s_{12})). \label{eq:regionb}
\end{eqnarray}
\end{theorem}

The proof of Theorem \ref{thm:region} is presented in Section \ref{sec:regionp}.

\subsection{Discrete Memoryless Keys}\label{sec:dmc}
The results for secure unicast and multicast with combinatorial keys generalize to discrete memoryless keys and are presented in the following theorem.

\begin{theorem}\label{thm:dmc}
{\normalfont[Secure Unicast and Multicast under Discrete Memoryless Keys]}
For the secure unicast problem (to the first of $K$ receivers), the capacity and the minimum broadcast bandwidth for capacity achieving schemes are
\begin{eqnarray}
C = \beta^*(C) = \min_{e \in [2:K]} H(z_1 | z_e).
\end{eqnarray}
For the secure multicast problem (to the first $K-1$ of $K$ receivers), the capacity is 
\begin{eqnarray}
C = \min_{q\in[1:K-1]} H(z_q | z_K).
\end{eqnarray}
\end{theorem}

The converse proof of Theorem \ref{thm:dmc} is identical to that under the combinatorial key setting.
The achievability proof of Theorem \ref{thm:dmc} is presented in Section \ref{sec:dmcp}. We give an intuitive overview here. 
First, consider secure unicast. Based on $z_1$, we wish to generate a key that is secure to any eavesdropping Receiver $e \in [2:K]$. With a discrete memoryless source, we can use random binning (whose mapping does not depend on $z_2, \cdots, z_K$) to obtain $H(z_1 | z_e)L + o(L)$ secure bits over $L$ key blocks. This step is well known and is typically referred to as privacy amplification \cite{Sudan_Tyagi_Watanabe} (refer to Lemma \ref{lemma:pa} in Section \ref{sec:dmcp} for a technical description). Given these secure bits, the rate value of the capacity is easily achieved by one-time pad coding.  
Second, consider secure multicast, which is similar, but with an additional step of communication for omniscience \cite{Csiszar_Narayan} (well known as well). This is implemented as follows. We assume the key $Z_K$ known by the eavesdropping Receiver $K$ is globally known (e.g., the transmitter may broadcast $Z_K$ to everyone). Next we wish to make the qualified receivers $1$ to $K-1$ all know $Z_1, \cdots, Z_{K-1}$ (i.e., common randomness). To this end, by Slepian Wolf coding \cite{Slepian_Wolf} (random binning), the transmitter needs to broadcast $\max_{q\in[1:K-1]} H(z_1, \cdots, z_{K-1} | z_q, z_K)L + o(L)$ bits over $L$ key blocks and note that these bits are available to the eavesdropping Receiver $K$ as well. After this communication for omniscience step, the qualified receivers all know $Z_1, \cdots, Z_{K-1}$ so that from privacy amplification (under eavesdropped public communication), they can agree on a key of size $(H(z_1, \cdots, z_{K-1}|z_K) - \max_{q\in[1:K-1]}  H(z_1, \cdots, z_{K-1} | z_q, z_K))L + o(L) = \min_{q\in[1:K-1]} H(z_q | z_K) L + o(L)$ bits that are almost unknown to the eavesdropping Receiver $K$ (i.e., the conditional entropy subtracts the amount of leaked communication). Equipped with these secure key bits, the desired rate can be easily achieved with one-time pad coding.

\begin{remark}
Similar to combinatorial secure multicast (see Remark \ref{remark:multicastcom}), the minimum broadcast bandwidth of secure multicast under the discrete memoryless key setting is an open problem. In particular, the step of communication for omniscience is not necessary and might cause additional broadcast bandwidth (this statement is also true for the key agreement problem \cite{Csiszar_Narayan}).
\end{remark}

\section{Related Work}\label{sec:prior}
The elemental problem of secure groupcast has interesting connections to several problems that have been studied in prior work and this section is devoted to the discussion of these connections. Due to space limits, we will focus on the connections to secure groupcast and leave further details to the references cited.

\subsubsection*{Secret Key Agreement (Generation)} 
In the problem of key agreement \cite{Maurer_Key, Ahlswede_Csiszar_CR, Csiszar_Narayan, Gohari_Anantharam, Chan_Zheng}, multiple terminals observing correlated sources wish to agree on a common key through public communication and it is required that an eavesdropper learns nothing about the key from public communication. 

Secret key agreement provides a natural achievable scheme for secure groupcast, where the legitimate receivers first agree on a secret key that is not known to the eavesdropping receivers (with the help of the transmitter and the noiseless broadcast channel). Then the secret key can be used to encrypt the desired message. This idea has been used in Section \ref{sec:dmc}. Unfortunately, secret key agreement is only understood when there is a {\em single} eavesdropper \cite{Csiszar_Narayan} but in secure groupcast, we have {\em multiple} eavesdroppers, each with a different view of the source. Also, for key agreement, only the maximum key rate (corresponding to the groupcast rate in secure groupcast) is known and the communication cost (corresponding to the broadcast bandwidth in secure groupcast) remains open in general \cite{Csiszar_Narayan}. Lastly, key agreement does not appear necessary for secure groupcast.


\subsubsection*{Latent Capacity Region} 
The latent capacity region of broadcast channels \cite{Grokop_Tse, Tian_Latent, Salimi_Liu_Cui} studies the implication of a rate tuple of common messages for various subsets of receivers being achievable, i.e., how can the achievable rates of certain group of receivers be exchanged for those of other groups of receivers? This interesting open problem is conceptually related to combinatorial secure groupcast, where from a rate exchange perspective, we are asking how to exchange various key variables shared by subsets of receivers to a common message for the group of desired receivers. However, latent capacity region has no security constraint and the required techniques in achievability and converse appear different.

\subsubsection*{Secure Broadcasting} 
How to send messages securely over a broadcast channel has been studied along the line of Wyner's wiretap channel \cite{Wyner_Wiretap}, and its generalizations to confidential messages (see e.g., \cite{Csiszar_Korner, Liu_Maric_Spasojevic_Yates}), and secure broadcasting over wireless channels (see e.g., \cite{Khisti_Tchamkerten_Wornell, Ekrem_Ulukus}). The enabler of secure communication in this line of work is that different receivers experience different channels, i.e., the channel itself has relative advantage to be exploited. In contrast, in secure groupcast every receiver sees the same noiseless broadcast channel and relative advantage comes from the shared keys. Notably, a recent work has studied a model (with a few users and a simple key structure) where both shared keys and discrete memoryless broadcast channels are simultaneously present \cite{Schaefer_Khisti_Poor}.

\subsubsection*{Secure (Private) Index Coding} 
Index coding \cite{Yossef_Birk_Jayram_Kol_Trans} is a canonical problem that studies how to efficiently broadcast under various side information at the receiver side with a noiseless broadcast channel. There are several variants of index coding that include security constraints (see e.g., \cite{Dau_Skachek_Chee_Security, Mojahedian_Aref_Gohari, Narayanan_Prabhakaran_Ravietal}) with and without shared keys and with and without external eavesdroppers. The main focus of index coding works is on the side information structure and its interplay with multiple desired messages. Secure groupcast highlights the shared key structure and its influence on message rate and broadcast bandwidth.

\section{Proofs}
\subsection{Proof of Theorem \ref{thm:msg}: Converse on $R$} \label{sec:msg}
Consider any qualified Receiver $q \in [1:N]$ and any eavesdropping Receiver $e \in [N+1:K]$. 
\begin{eqnarray}
RL &\overset{(\ref{rate})}{=}& H(W)  \\
&\overset{(\ref{wzind})}{=}& H(W | Z_e) \\
&\overset{(\ref{corr})}{=}& I(W; X, Z_q | Z_e) + o(L) \\
&\overset{(\ref{sec})}{=}& I(W; Z_q | X, Z_e) + o(L) \\
&\leq& H(Z_q | Z_e) + o(L) \\
&=& L H(z_q | z_e) + o(L). \label{eq:msgf}
\end{eqnarray}
Normalizing (\ref{eq:msgf}) by $L$ and letting $L \rightarrow \infty$, we have the desired converse bound $R \leq H(z_q | z_e)$.

\subsection{Proof of Theorem \ref{thm:bw}: Converse on $\beta(R)$}\label{sec:bw}
To simplify the notations, we set $\mathcal{Q} = \{1,2,\cdots, Q\} \subset [1:N]$, which has no loss of generality. Let us start with a useful lemma.
\begin{lemma}\label{lemma:msgo}
For any $q \in [1:Q-1]$, we have
\begin{eqnarray}
I(X; Z_{q+1} | Z_1, \cdots, Z_{q}, U_e, W) \geq H(W) - I(Z_1, \cdots, Z_{q}; Z_{q+1} | U_e) + o(L). \label{eq:msgo}
\end{eqnarray}
\end{lemma}

{\it Proof:}
\begin{eqnarray}
&& I(X; Z_{q+1} | Z_1, \cdots, Z_{q}, U_e, W) \notag\\
&\overset{(\ref{wzind})}{=}& I(X, W; Z_{q+1} | Z_1, \cdots, Z_{q}, U_e) \\
&=&  I(X, W,  Z_1, \cdots, Z_{q}; Z_{q+1} | U_e)  -  I(Z_1, \cdots, Z_{q}; Z_{q+1} | U_e) \\
&\geq& I(W ; Z_{q+1} | U_e, X)  -  I(Z_1, \cdots, Z_{q}; Z_{q+1} | U_e) \\
&\overset{(\ref{corr})}{=}& H(W | U_e, X) -  I(Z_1, \cdots, Z_{q}; Z_{q+1} | U_e)  + o(L) \\
&\overset{(\ref{sec})}{=}& H(W | U_e) -  I(Z_1, \cdots, Z_{q}; Z_{q+1} | U_e)  + o(L) \\
&\overset{(\ref{wzind})}{=}& H(W) - I(Z_1, \cdots, Z_{q}; Z_{q+1} | U_e)  + o(L).
\end{eqnarray}
\hfill\QED

Next, we apply Lemma \ref{lemma:msgo} to decompose the term $I(X; W, Z_1, \cdots, Z_Q | U_e)$.
\begin{eqnarray}
&& I(X; W, Z_1, \cdots, Z_Q | U_e) \notag\\
&=& I(X; W, Z_1 | U_e) + \sum_{q=1}^{Q-1} I(X; Z_{q+1} | Z_1, \cdots, Z_{q}, U_e, W) \\
&\overset{(\ref{eq:msgo})}{\geq}& I(X; W | Z_1, U_e) + \sum_{q=1}^{Q-1}  \Big( H(W) - I(Z_1, \cdots, Z_{q}; Z_{q+1} | U_e)  + o(L) \Big)\\
&\overset{(\ref{wzind})}{=}& I(X, Z_1, U_e; W) + \sum_{q=1}^{Q-1}  \Big( H(W) - I(Z_1, \cdots, Z_{q}; Z_{q+1} | U_e)  + o(L) \Big) \\
&\overset{(\ref{corr})}{=}& H(W) + o(L) + \sum_{q=1}^{Q-1}  \Big( H(W) - I(Z_1, \cdots, Z_{q}; Z_{q+1} | U_e)  + o(L) \Big) \\
&=& Q H(W) - \sum_{q=1}^{Q-1} I(Z_1, \cdots, Z_{q}; Z_{q+1} | U_e) + o(L) . \label{eq:msg1}
\end{eqnarray}

Finally, note that
\begin{eqnarray}
\beta(R) L &\overset{(\ref{bw})}{=}&  L_X \log_2 p \\
&\geq& H(X) \\
&\geq& H(X|U_e) \\ 
&\geq& I(X; W, Z_1, \cdots, Z_Q | U_e) . \label{eq:msg2}
\end{eqnarray}

Combining (\ref{eq:msg1}), (\ref{eq:msg2}), we have
\begin{eqnarray}
\beta(R) L \geq Q H(W) - \sum_{q=1}^{Q-1} I(Z_1, \cdots, Z_{q}; Z_{q+1} | U_e) + o(L). \label{eq:msg3}
\end{eqnarray}
Normalizing (\ref{eq:msg3}) by $L$ and letting $L \rightarrow \infty$, we have the desired converse bound on $\beta(R)$.

\subsection{Proof of Theorem \ref{thm:unicast}: Secure Unicast}\label{sec:unicast}
The converse bounds for the capacity and the broadcast bandwidth follow immediately from Theorem \ref{thm:msg} and Theorem \ref{thm:bw}, respectively. The achievable scheme is presented as follows. 

We show that rate $R = \min_{e \in [2:K]} \sum_{\mathcal{U} \subset [1:K]:  1 \in \mathcal{U}, e \notin \mathcal{U}} H(s_{\mathcal{U}})$ and broadcast bandwidth $\beta(R) = R$ are achievable. Set $L_W = R/\log_2 p$ and $L=1$. Set the field size $p$ to be the least prime power such that $p \geq L_W + \sum_{\mathcal{U} \subset [1:K]:  1 \in \mathcal{U}} L_{\mathcal{U}}$.
\begin{eqnarray}
X &=& W + \sum_{\mathcal{U} \subset [1:K]:  1 \in \mathcal{U}} {\bf V}_{\mathcal{U}} s_{\mathcal{U}} = W + \underbrace{\Big[{\bf V}_{1} ~ {\bf V}_{12}~\cdots~{\bf V}_{1:K} \Big]}_{\triangleq {\bf V}}
\left[ \begin{array}{c}
s_{1} \\
s_{12} \\
\vdots\\
s_{1:K}
\end{array}
\right] \label{eq:unis}
\end{eqnarray}
where $X, W \in \mathbb{F}_p^{L_W \times 1}$, ${\bf V}_{\mathcal{U}} \in \mathbb{F}_p^{L_W \times L_{\mathcal{U}}}$, $s_{\mathcal{U}}  \in \mathbb{F}_p^{L_{\mathcal{U}} \times 1}$ and ${\bf V}$ is chosen as a full-rank Cauchy matrix of dimension $L_W \times \sum_{\mathcal{U} \subset [1:K]:  1 \in \mathcal{U}} L_{\mathcal{U}}$ such that the element in $i$-th row and $j$-th column is given by
\begin{eqnarray}
{\bf V}(i,j) = \frac{1}{a_i - b_j}, ~~\mbox{$a_i, b_j$ are distinct elements over $\mathbb{F}_p$.}
\end{eqnarray}

The correctness constraint (\ref{corr}) is trivially satisfied. We verify the security constraint (\ref{sec}). Consider any eavesdropping Receiver $e\in [2:K]$.
\begin{eqnarray}
I(W; X, Z_e) &\overset{(\ref{wzind})(\ref{eq:unis})}{=}& I(W; W + \sum_{\mathcal{U} \subset [1:K]:  1 \in \mathcal{U}} {\bf V}_{\mathcal{U}} s_{\mathcal{U}}  | z_e) \\
&=& I\Big( W; W + \sum_{\mathcal{U} \subset [1:K]:  1 \in \mathcal{U}, e\notin \mathcal{U}} {\bf V}_{\mathcal{U}} s_{\mathcal{U}}  | (s_{\mathcal{U}} : e \in \mathcal{U}) \Big) \\
&\overset{(\ref{wzind})(\ref{combi})}{=}& I\Big( W; W + \sum_{\mathcal{U} \subset [1:K]:  1 \in \mathcal{U}, e\notin \mathcal{U}} {\bf V}_{\mathcal{U}} s_{\mathcal{U}}  \Big) \\
&\overset{(\ref{wzind})}{=}& H\Big(W + \sum_{\mathcal{U} \subset [1:K]:  1 \in \mathcal{U}, e\notin \mathcal{U}} {\bf V}_{\mathcal{U}} s_{\mathcal{U}}  \Big) - H\Big(\sum_{\mathcal{U} \subset [1:K]:  1 \in \mathcal{U}, e\notin \mathcal{U}} {\bf V}_{\mathcal{U}} s_{\mathcal{U}}  \Big) \\
&\leq& L_W \log_2 p - L_W \log_2 p = 0
\end{eqnarray}
where in the last step, the first term of $L_W \log_2 p$ follows from the fact that the vector has $L_W$ symbols from $\mathbb{F}_p$ and the second term of $-L_W \log_2 p$ follows from the fact that the sub-matrix $[{\bf V}_{\mathcal{U}}: \mathcal{U} \subset [1:K], 1 \in \mathcal{U}, e\notin \mathcal{U}]$ of the Cauchy matrix ${\bf V}$ has rank $L_W$ (as it has at least $L_W$ columns and exactly $L_W$ rows) and $s_{\mathcal{U}}$ are i.i.d. uniform symbols.

Finally, the rate and broadcast bandwidth achieved match the converse such that the proof of Theorem \ref{thm:unicast} is complete.

\subsection{Proof of Theorem \ref{thm:multicast}: Secure Multicast}\label{sec:multicast}
We first consider the capacity of combinatorial secure multicast.
The converse follows directly from Theorem \ref{thm:msg} and we consider the achievability.

We show that rate $R = \min_{q\in[1:K-1]} \sum_{\mathcal{U} \subset [1:K-1]:q\in\mathcal{U}} H(s_{\mathcal{U}})$ is achievable. 
Set $L_W = R/\log_2 p$ and $L=1$. Set the field size $p$ to be the least prime power such that $p \geq L_W + \sum_{\mathcal{U} \subset [1:K-1]} L_{\mathcal{U}}$. The transmit signal has $2^{K-1}-1$ row blocks, each corresponding to a subset of $[1:K-1]$.
\begin{eqnarray}
X &=& [X_{1}; X_2; \cdots; X_{\mathcal{U}}; \cdots; X_{1:K-1}] \notag \\
X_{\mathcal{U}} &=& {\bf V}_{\mathcal{U}}W + s_{\mathcal{U}}, \forall \mathcal{U} \subset [1:K-1]
\label{eq:multicast}
\end{eqnarray}
where $X_{\mathcal{U}} \in \mathbb{F}_p^{L_\mathcal{U} \times 1}, {\bf V}_{\mathcal{U}} \in \mathbb{F}_p^{L_\mathcal{U} \times L_W}, W\in\mathbb{F}_p^{L_W\times 1}, s_{\mathcal{U}} \in \mathbb{F}_p^{L_{\mathcal{U}}\times1}$ and the precoding matrices ${\bf V}_{\mathcal{U}}$ are sub-matrices of a Cauchy matrix, set as follows.
\begin{eqnarray}
{\bf V} &=& [{\bf V}_{1}; {\bf V}_2; \cdots; {\bf V}_{\mathcal{U}}; \cdots; {\bf V}_{1:K-1}]_{\sum_{\mathcal{U} \subset [1:K-1]} L_{\mathcal{U}} \times L_W},\\
{\bf V}(i,j) &=& \frac{1}{a_i - b_j}, ~~\mbox{$a_i, b_j$ are distinct elements over $\mathbb{F}_p$.}
\end{eqnarray}

Security is guaranteed because the keys known to the eavesdropping Receiver $K$ do not appear in the transmit signal and the $s_{\mathcal{U}}$ variables are independent. Correctness constraint is satisfied because each qualified Receiver $q\in[1:K-1]$ can recover at least $L_W$ linear combinations of the message symbols $W$, i.e., $({\bf V}_{\mathcal{U}}W : q \in \mathcal{U} \subset[1:K-1])$, from which $W$ can be decoded as any sub-matrix of a full-rank Cauchy matrix has full rank.

Next we proceed to the minimum broadcast bandwidth of combinatorial secure multicast. The broadcast bandwidth achieved by the scheme above is $\beta(C) = \sum_{\mathcal{U} \subset [1:K-1]} H(s_{\mathcal{U}})$, which is optimal when $H(z_1|z_K) = \cdots = H(z_{K-1}|z_K)$. This follows from Theorem \ref{thm:bw}, where we set $\mathcal{Q} = [1:K-1]$ and $u_e = z_K$.
\begin{eqnarray}
\beta(C) &\geq& (K-1) C - (\sum_{q=1}^{K-1} H(z_q | z_K) - H(z_1, \cdots, z_{K-1} | z_K))\\
&=& H(z_1, \cdots, z_{K-1} | z_K) = \sum_{\mathcal{U} \subset [1:K-1]} H(s_{\mathcal{U}}) \label{eq:mulc}
\end{eqnarray}
where (\ref{eq:mulc}) follows from the fact that $C = H(z_1 | z_K) = \cdots = H(z_{K-1} | z_K)$. The proof when $K=4$ is more involved as we need to improve the above achievable scheme, and is presented next.

\subsubsection{$\beta^*(C)$ when $K = 4$} \label{sec:mulbw}
First, we consider the converse. Note that in this case, $C = H(z_1|z_4) $. From (\ref{eq:multicast4}), we need two converse bounds. 
The first one is obtained by setting $\mathcal{Q} = \{1,2\}, u_e = z_4$ in Theorem \ref{thm:bw},
\begin{eqnarray}
\beta(C) &\geq& 2C - I(z_1; z_2|z_4) = 2H(z_1|z_4) - I(z_1; z_2|z_4)  \\
&=& 2H(s_1) + 2H(s_{12}) + 2H(s_{13}) + 2H(s_{123}) - (H(s_{12}) + H(s_{123}))\\
&=& 2H(s_1) + H(s_{12}) + 2H(s_{13}) + H(s_{123}). \label{eq:mulbw1}
\end{eqnarray}
The second one is obtained by setting $\mathcal{Q} = \{1,2,3\}, u_e = z_4$ in Theorem \ref{thm:bw},
\begin{eqnarray}
\beta(C) &\geq& 3C - (I(z_1; z_2|z_4) + I(z_1, z_2; z_3| z_4)) \\
&=& 3H(s_1) + 3H(s_{12}) + 3H(s_{13}) + 3H(s_{123}) - (H(s_{12}) +  H(s_{13}) + H(s_{23}) + 2H(s_{123})) \notag\\
&=& 3H(s_1) + 2H(s_{12}) + 2H(s_{13}) - H(s_{23}) + H(s_{123}). \label{eq:mulbw2}
\end{eqnarray}

Second, we consider the achievability, where we need to adjust the size of the keys used in (\ref{eq:multicast}) depending on the key configuration.
We present the scheme that achieves rate $R = H(z_1|z_4) = H(s_1) + H(s_{12}) + H(s_{13}) + H(s_{123})$. 
Set $L_W = R/\log_2 p$ and $L=1$. 

We have $3$ cases depending on the relationship between $H(s_{23}), H(s_{1})+H(s_{13}), H(s_1)+H(s_{12})$.  
For each case, set the field size $p$ to be the least prime power such that $p \geq L_X + L_W$. The transmit signal has $7$ row blocks and the $4$ blocks $X_1, X_{12}, X_{13}, X_{123}$ (corresponding to the keys known to Receiver 1) are the same for all $3$ cases, where all key symbols are used.
\begin{eqnarray}
X &=& [X_{1}; X_2; X_3; X_{12}; X_{13}; X_{23}; X_{123}] \notag \\
X_1 &=& {\bf V}_1 W + s_1, X_{12} = {\bf V}_{12} W + s_{12}, X_{13} = {\bf V}_{13} W + s_{13}, X_{123} = {\bf V}_{123} W + s_{123}
\end{eqnarray}
where the sizes of the matrices and vectors above are the same as before (see (\ref{eq:multicast})).
Note that now Receiver 1 can achieve rate $R$ and in the remaining proof, we only need to consider Receiver 2 and Receiver 3.
The remaining blocks $X_2, X_3, X_{23}$ are designed for each case separately, where not all the key symbols may be used. Note that $H(s_{12}) \leq H(s_{13})$, i.e., $L_{12} \leq L_{13}$.
\begin{enumerate}
\item[] Case 1. $H(s_{23}) \geq H(s_1) + H(s_{13})$.
\begin{eqnarray}
X_2 = (), X_3 = (), X_{23} = {\bf V}_{23}^{[1]} W + s_{23}^{[1]}
\end{eqnarray}
where $s_{23}^{[1]} \in \mathbb{F}_p^{(L_1+L_{13}) \times 1}$ is comprised of the first $L_1+L_{13}$ symbols from $s_{23}$ (which has more symbols, i.e., $L_{23} \geq L_1+L_{13}$) and ${\bf V}_{23}^{[1]} \in \mathbb{F}_p^{(L_1+L_{13})\times L_W}$. We verify that the number of linear combinations of the message decodable by Receiver 2 and Receiver 3 is no less than $L_W$.
\begin{eqnarray}
\mbox{For Receiver 2:}~|X_{12}| + |X_{23}| + |X_{123}| = L_{12} + (L_{1} + L_{13}) + L_{123} = L_W, \\
\mbox{For Receiver 3:}~|X_{13}| + |X_{23}| + |X_{123}| = L_{13} + (L_{1} + L_{13}) + L_{123} \geq L_W.
\end{eqnarray}
The transmit signal size is
\begin{eqnarray}
L_X = |X_{1}| + |X_{12}| + |X_{13}| + |X_{123}| + |X_{23}| = L_1 + L_{12} + L_{13} + L_{123} + (L_1 + L_{13})
\end{eqnarray} 
which matches the converse bound (\ref{eq:mulbw1}) for broadcast bandwidth. The other cases are similar.
\item[] Case 2. $H(s_1) + H(s_{12}) \leq H(s_{23}) \leq H(s_1) + H(s_{13})$.
\begin{eqnarray}
X_2 = {\bf V}_2^{[2]} W + s_{2}^{[2]}, X_3 = (), X_{23} = {\bf V}_{23} W + s_{23}
\end{eqnarray}
where $s_{2}^{[2]} \in \mathbb{F}_p^{(L_1+L_{13}-L_{23}) \times 1}$ is comprised of the first $L_1+L_{13} - L_{23}$ symbols from $s_{2}$ (which has more symbols, i.e., $L_{2} \geq L_1+L_{13} - L_{23}$ because $H(z_1|z_4) \leq H(z_2|z_4)$) and ${\bf V}_{2}^{[2]} \in \mathbb{F}_p^{(L_1+L_{13} - L_{23})\times L_W}$. We verify that the number of linear combinations of the message decodable by Receiver 2 and Receiver 3 is no less than $L_W$.
\begin{eqnarray}
&& \mbox{For Receiver 2:}~|X_2| + |X_{12}| + |X_{23}| + |X_{123}| = (L_{1} + L_{13} - L_{23}) + L_{12} + L_{23} + L_{123}  \notag \\
&& ~~~~~~~~~~~~~~~~~~~~= L_{1} + L_{12} + L_{13} + L_{123} = L_W, \\
&& \mbox{For Receiver 3:}~|X_{13}| + |X_{23}| + |X_{123}| = L_{13} + L_{23} + L_{123} \geq L_W. 
\end{eqnarray}
The transmit signal size is
\begin{eqnarray}
L_X &=& |X_{1}| + |X_{12}| + |X_{13}| + |X_{123}| + |X_2| + |X_{23}| \\
&=& L_1 + L_{12} + L_{13} + L_{123} + (L_1 + L_{13} - L_{23}) + L_{23}
\end{eqnarray} 
which matches the converse bound (\ref{eq:mulbw1}) for broadcast bandwidth.
\item[] Case 3. $H(s_{23}) \leq H(s_1) + H(s_{12})$.
\begin{eqnarray}
&&X_2 = {\bf V}_2^{[3]} W + s_{2}^{[3]}, X_3 = {\bf V}_3^{[3]} W + s_{3}^{[3]}, X_{23} = {\bf V}_{23} W + s_{23} \\
&& s_{2}^{[3]} \in \mathbb{F}_p^{(L_1+L_{13}-L_{23}) \times 1}, {\bf V}_{2}^{[3]} \in \mathbb{F}_p^{(L_1+L_{13} - L_{23})\times L_W} \\
&& s_{3}^{[3]} \in \mathbb{F}_p^{(L_1+L_{12}-L_{23}) \times 1}, {\bf V}_{3}^{[3]} \in \mathbb{F}_p^{(L_1+L_{12} - L_{23})\times L_W}
\end{eqnarray}
where $L_{2} \geq L_1+L_{13} - L_{23}$ and $L_3 \geq L_1+L_{12}-L_{23}$ because $H(z_1|z_4) \leq \min( H(z_2|z_4), H(z_3|z_4))$. We verify that the number of linear combinations of the message decodable by Receiver 2 and Receiver 3 is no less than $L_W$.
\begin{eqnarray}
&& \mbox{For Receiver 2:}~|X_2| + |X_{12}| + |X_{23}| + |X_{123}| = (L_{1} + L_{13} - L_{23}) + L_{12} + L_{23} + L_{123}  \notag \\
&& ~~~~~~~~~~~~~~~~~~~~= L_{1} + L_{12} + L_{13} + L_{123} = L_W, \\
&& \mbox{For Receiver 3:}~|X_3| + |X_{13}| + |X_{23}| + |X_{123}| = (L_{1} + L_{12} - L_{23}) + L_{13} + L_{23} + L_{123} \notag\\
&& ~~~~~~~~~~~~~~~~~~~~= L_{1} + L_{12} + L_{13} + L_{123} = L_W.
\end{eqnarray}
The transmit signal size is
\begin{eqnarray}
L_X &=& |X_{1}| + |X_{12}| + |X_{13}| + |X_{123}| + |X_2| + |X_3| + |X_{23}| \\
&=& L_1 + L_{12} + L_{13} + L_{123} + (L_1 + L_{13} - L_{23}) + (L_1 + L_{12} - L_{23}) + L_{23}
\end{eqnarray} 
which matches the converse bound (\ref{eq:mulbw2}) for broadcast bandwidth.
\end{enumerate}

After the sizes are specified, the remaining proof is the same as that presented above, where we set ${\bf V}$ as a full-rank Cauchy matrix and the correctness and security constraints are satisfied. The proof of $\beta^*(C)$ when $K=4$ is thus complete.


\subsection{Proof of Theorem \ref{thm:42}: The $N=2, K=4$ Case}\label{sec:42}
The converse of rate and broadcast bandwidth follows from Theorem \ref{thm:msg} and Theorem \ref{thm:bw}, and we present the achievable scheme now. Set $p=2$ (i.e., binary field) and $L=1$ (i.e., one key block). The idea is to decompose every instance into multiple component sub-networks, where the basic components are listed in the following figure.

\begin{figure}[h]
\begin{center}
\includegraphics[width=7 in]{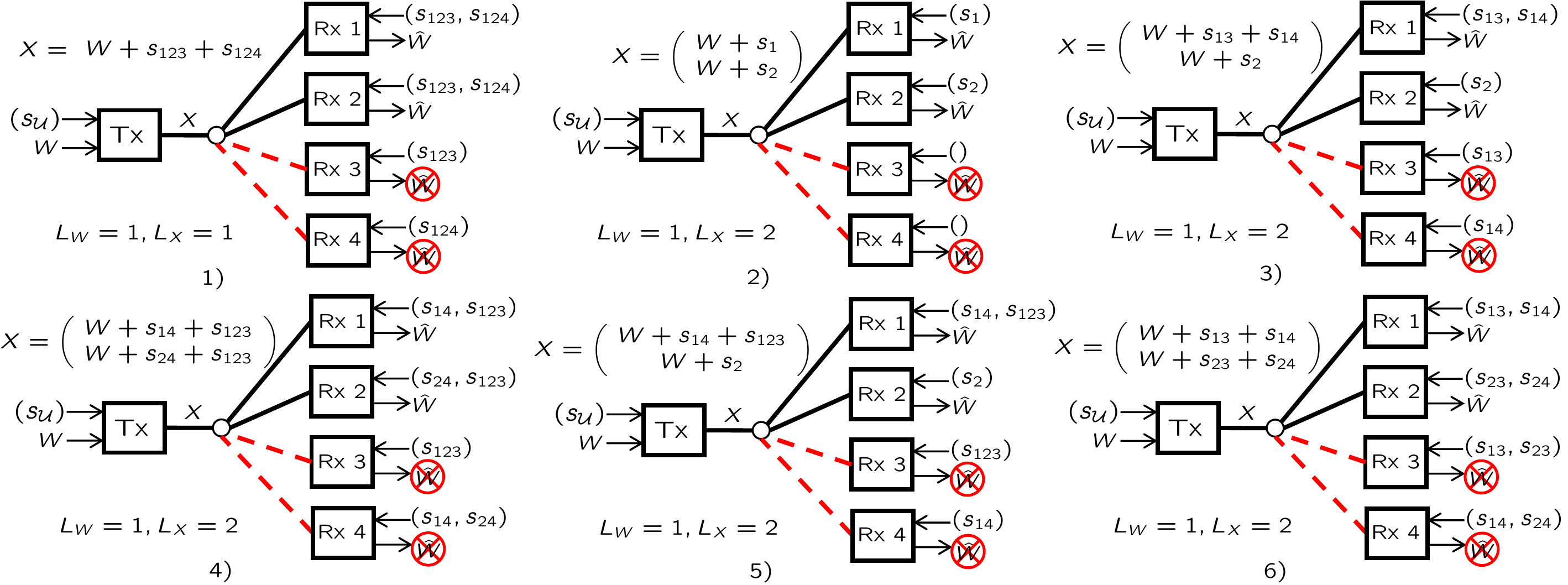}
\caption{\small The $6$ basic components with (correct and secure) achievable schemes.}
\label{fig:ex42}
\end{center}
\end{figure}

Next, we divide the problem instance into multiple cases, where each case requires a different decomposition. As $s_{12}$ is only known to qualified Receiver $1$ and Receiver $2$, we can easily use $s_{12}$ to achieve rate $H(s_{12})$ with broadcast bandwidth $H(s_{12})$, by one-time pad. 
Without loss of generality, we assume $H(s_1)\leq H(s_2), H(s_{124}) \leq H(s_{123})$. As such, for all cases we invoke Component $1$ $H(s_{124})$ times (i.e., use $H(s_{124})$ bits of $s_{123}, s_{124}$) and Component $2$ $H(s_{1})$ times (use $H(s_1)$ bits of $s_1, s_2$). We write this succinctly as
\begin{eqnarray}
H(s_{124}) \times \mbox{~Cmp $1$~} + H(s_{1}) \times \mbox{~Cmp $2$}. \label{eq:421}
\end{eqnarray}
We proceed next depending on the key configurations. Specifically, all cases are divided as follows.
\begin{enumerate}
\item[] Case 1. $H(s_2) - H(s_1) \geq \min(H(s_{13}), H(s_{14}))$. 
We further invoke
\begin{eqnarray}
\min(H(s_{13}), H(s_{14})) \times \mbox{~Cmp $3$~} \label{eq:422}
\end{eqnarray}
where we can employ the scheme in Component $3$ a number of $\min(H(s_{13}), H(s_{14}))$ times because we have $H(s_2) - H(s_1)$ bits left of $s_2$ and $H(s_2) - H(s_1) \geq \min(H(s_{13}), H(s_{14}))$.
The remaining steps need further division.
\subitem Case 1.1. $H(s_{14}) \leq H(s_{13})$. No further action is needed. Tracing back, we have invoked one-time pad of $s_{12}$, (\ref{eq:421}), and (\ref{eq:422}). Therefore we have achieved 
\begin{eqnarray}
R = H(s_{12}) + H(s_{124}) + H(s_{1}) + H(s_{14}), ~\beta(R) = H(s_{12}) + H(s_{124}) + 2H(s_{1}) + 2H(s_{14}) \notag
\end{eqnarray}
which match the converse (\ref{eq:42msg}) and (\ref{eq:42bw}). Note that the converse bounds are minimum or maximum of several terms and it suffices to show the achievability of one term.
\subitem Case 1.2. $H(s_{14}) \geq H(s_{13})$. We need to further consider the following cases.
\subsubitem Case 1.2.1. $\min(H(s_{14}) - H(s_{13}), H(s_{123}) - H(s_{124}), H(s_{24})) = H(s_{14}) - H(s_{13})$. We further invoke
\begin{eqnarray}
(H(s_{14}) - H(s_{13})) \times \mbox{~Cmp $4$~}
\end{eqnarray}
and the description of the scheme is complete for this case. We trace back and find that
\begin{eqnarray}
R = H(s_{12}) + H(s_{124}) + H(s_{1}) + H(s_{14}), ~\beta(R) = H(s_{12}) + H(s_{124}) + 2H(s_{1}) + 2H(s_{14}) \notag
\end{eqnarray}
are achieved and they are optimal as the formulas match the converse.
\subsubitem Case 1.2.2. $\min(H(s_{14}) - H(s_{13}), H(s_{123}) - H(s_{124}), H(s_{24})) =  H(s_{123}) - H(s_{124})$. We further invoke
\begin{eqnarray}
(H(s_{123}) - H(s_{124})) \times \mbox{~Cmp $4$~}
\end{eqnarray}
and the description of the scheme is complete for this case. We trace back and find that
\begin{eqnarray}
R &=& H(s_{12}) + H(s_{1}) + H(s_{13}) + H(s_{123}), \\
\beta(R) &=& H(s_{12}) - H(s_{124}) + 2H(s_{1}) + 2H(s_{13}) + 2 H(s_{123})
\end{eqnarray}
are achieved and they are optimal as the formulas match the converse.
\subsubitem Case 1.2.3. $\min(H(s_{14}) - H(s_{13}), H(s_{123}) - H(s_{124}), H(s_{24})) = H(s_{24})$. We further invoke 
\begin{eqnarray}
H(s_{24}) \times \mbox{~Cmp $4$~}
\end{eqnarray}
and need to consider the following cases. 
\begin{enumerate}
\item[] ~~~~Case 1.2.3.1. $\min(H(s_{2}) - H(s_{1}) - H(s_{13}), H(s_{14}) - H(s_{13}) - H(s_{24}), H(s_{123}) - H(s_{124}) - H(s_{24}) ) = H(s_{2}) - H(s_{1}) - H(s_{13})$. We further invoke
\begin{eqnarray}
&& \big( H(s_{2}) - H(s_{1}) - H(s_{13}) \big) \times \mbox{~Cmp $5$~ such that overall} \\
&& R = H(s_{12}) + H(s_{124}) + H(s_{24}) + H(s_2), \\
&& \beta(R) = H(s_{12}) + H(s_{124}) + 2H(s_{24}) + 2H(s_{2}).
\end{eqnarray}
\item[] ~~~~Case 1.2.3.2. $\min(H(s_{2}) - H(s_{1}) - H(s_{13}), H(s_{14}) - H(s_{13}) - H(s_{24}), H(s_{123}) - H(s_{124}) - H(s_{24}) ) = H(s_{14}) - H(s_{13}) - H(s_{24})$. We further invoke
\begin{eqnarray}
&& \big( H(s_{14}) - H(s_{13}) - H(s_{24}) \big) \times \mbox{~Cmp $5$~ such that overall} \\
&& R = H(s_{12}) + H(s_{124}) + H(s_1) +  H(s_{14}), \\
&& \beta(R) = H(s_{12}) + H(s_{124}) + 2H(s_1) +  2H(s_{14}).
\end{eqnarray}
\item[] ~~~~Case 1.2.3.3. $\min(H(s_{2}) - H(s_{1}) - H(s_{13}), H(s_{14}) - H(s_{13}) - H(s_{24}), H(s_{123}) - H(s_{124}) - H(s_{24}) ) = H(s_{123}) - H(s_{124}) - H(s_{24})$. 
We further invoke
\begin{eqnarray}
&& \big( H(s_{123}) - H(s_{124}) - H(s_{24}) \big) \times \mbox{~Cmp $5$~ such that overall} \\
&& R = H(s_{12}) + H(s_{1}) +H(s_{13}) + H(s_{123}), \\
&& \beta(R) = H(s_{12}) - H(s_{124}) + 2H(s_{1}) +2H(s_{13}) + 2H(s_{123}).
\end{eqnarray}
\end{enumerate}
\item[] Case 2. $H(s_2) - H(s_1) \leq \min(H(s_{13}), H(s_{14}))$. We further invoke
\begin{eqnarray}
\big(H(s_2) - H(s_1)\big) \times \mbox{~Cmp $3$~}
\end{eqnarray}
and consider the following cases.
\subitem Case 2.1. $\min(H(s_{24}), H(s_{14}) - H(s_{2}) + H(s_{1}), H(s_{123}) - H(s_{124})) = H(s_{24})$. We further invoke
\begin{eqnarray}
H(s_{24})  \times \mbox{~Cmp $4$~}
\end{eqnarray}
and the description of the scheme is complete for this case. We trace back and find that
\begin{eqnarray}
R &=& H(s_{12}) + H(s_{124}) + H(s_{2}) + H(s_{24}), \\
\beta(R) &=&  H(s_{12}) + H(s_{124}) + 2H(s_{2}) + 2H(s_{24})
\end{eqnarray}
are achieved and they are optimal as the formulas match the converse.
\subitem Case 2.2. $\min(H(s_{24}), H(s_{14}) - H(s_{2}) + H(s_{1}), H(s_{123}) - H(s_{124})) = H(s_{14}) - H(s_{2}) + H(s_{1})$. We further invoke
\begin{eqnarray}
\big( H(s_{14}) - H(s_{2}) + H(s_{1})\big)  \times \mbox{~Cmp $4$~}
\end{eqnarray}
and the description of the scheme is complete for this case. We trace back and find that
\begin{eqnarray}
R &=& H(s_{12}) + H(s_{124})  + H(s_{14}) + H(s_{1}), \\
\beta(R) &=&  H(s_{12}) + H(s_{124}) + 2H(s_{14}) + 2H(s_{1})
\end{eqnarray}
are achieved and they are optimal as the formulas match the converse.
\subitem Case 2.3. $\min(H(s_{24}), H(s_{14}) - H(s_{2}) + H(s_{1}), H(s_{123}) - H(s_{124})) = H(s_{123}) - H(s_{124})$. We further invoke
\begin{eqnarray}
\big( H(s_{123}) - H(s_{124}) \big)  \times \mbox{~Cmp $4$~}
\end{eqnarray}
and need to consider the following cases. 
\subsubitem Case 2.3.1. $\min(H(s_{14}) - H(s_{2}) + H(s_{1}) - H(s_{123}) + H(s_{124}), H(s_{13}) - H(s_{2}) + H(s_{1}),  H(s_{24}) - H(s_{123}) + H(s_{124}), H(s_{23})) = H(s_{14}) - H(s_{2}) + H(s_{1}) - H(s_{123}) + H(s_{124})$. We further invoke
\begin{eqnarray}
&& \big( H(s_{14}) - H(s_{2}) + H(s_{1}) - H(s_{123}) + H(s_{124}) \big) \times \mbox{~Cmp $6$~ such that overall} \\
&& R = H(s_{12}) + H(s_{124}) + H(s_{14}) + H(s_{1}), \\
&& \beta(R) = H(s_{12}) + H(s_{124}) +  2H(s_{14}) +  2H(s_1).
\end{eqnarray}
\subsubitem Case 2.3.2. $\min(H(s_{14}) - H(s_{2}) + H(s_{1}) - H(s_{123}) + H(s_{124}), H(s_{13}) - H(s_{2}) + H(s_{1}),  H(s_{24}) - H(s_{123}) + H(s_{124}), H(s_{23})) = H(s_{13}) - H(s_{2}) + H(s_{1})$. We further invoke
\begin{eqnarray}
&& \big( H(s_{13}) - H(s_{2}) + H(s_{1}) \big) \times \mbox{~Cmp $6$~ such that overall} \\
&& R = H(s_{12}) + H(s_{123}) + H(s_{13}) + H(s_{1}),\\
&& \beta(R) = H(s_{12}) - H(s_{124}) +  2 H(s_{123}) + 2H(s_{13}) + 2H(s_{1}).
\end{eqnarray}
\subsubitem Case 2.3.3. $\min(H(s_{14}) - H(s_{2}) + H(s_{1}) - H(s_{123}) + H(s_{124}), H(s_{13}) - H(s_{2}) + H(s_{1}),  H(s_{24}) - H(s_{123}) + H(s_{124}), H(s_{23})) = H(s_{24}) - H(s_{123}) + H(s_{124})$. We further invoke
\begin{eqnarray}
&& \big( H(s_{24}) - H(s_{123}) + H(s_{124}) \big) \times \mbox{~Cmp $6$~ such that overall} \\
&& R = H(s_{12}) + H(s_{124}) + H(s_{2}) + H(s_{24})\\
&& \beta(R) = H(s_{12}) + H(s_{124}) +  2 H(s_{2}) + 2H(s_{24})
\end{eqnarray}
\subsubitem Case 2.3.4. $\min(H(s_{14}) - H(s_{2}) + H(s_{1}) - H(s_{123}) + H(s_{124}), H(s_{13}) - H(s_{2}) + H(s_{1}),  H(s_{24}) - H(s_{123}) + H(s_{124}), H(s_{23})) = H(s_{23})$. We further invoke
\begin{eqnarray}
&& H(s_{23}) \times \mbox{~Cmp $6$~ such that overall} \\
&& R = H(s_{12}) + H(s_{2}) + H(s_{123}) + H(s_{23}),\\
&& \beta(R) = H(s_{12}) - H(s_{124}) +  2 H(s_{2}) + 2H(s_{123}) + 2H(s_{23}).
\end{eqnarray}
\end{enumerate}

\subsection{Proof of Theorem \ref{thm:sym}: The Symmetric Setting}\label{sec:sym}
The rate converse follows from Theorem \ref{thm:msg}, where among $u$-keys, any qualified Receiver $q \in [1:N]$ knows $\binom{K-2}{u-1}$ keys that are not known to any eavesdropping Receiver $e \in [N+1:K]$ because we may pick any $u-1$ receivers from any $K-2$ receivers other than Receiver $q$ and Receiver $e$ to form a $u$-key (note that Receiver $q$ must be included). The broadcast bandwidth converse follows from Theorem \ref{thm:bw}, where we set $\mathcal{Q} = [1:N]$, $u_e = z_K$ and obtain $\beta(C) \geq H(z_1,\cdots,z_N|z_K)$. Among $u$-keys, we have $\binom{K-1}{u} - \binom{K-N-1}{u}$ keys in the term $H(z_1,\cdots,z_N|z_K)$ because we pick $u$-keys from receivers $1$ to $K-1$ and need to remove the ones that are only known to receivers $N+1$ to $K-1$. To sum up for the converse part, we have proved that
\begin{eqnarray}
C \leq \sum_{u=1}^K \binom{K-2}{u-1}L^{[u]}\log_2 p, ~~\beta^*(C) \geq \sum_{u=1}^K \left( \binom{K-1}{u} - \binom{K-N-1}{u} \right) L^{[u]} \log_2 p.
\end{eqnarray}
We next show that the above rate and broadcast bandwidth are achievable. Similar to Example \ref{ex:sym}, we consider $u$-keys separately for different $u$ values and then combine the decomposed schemes using Lemma \ref{lemma:decompose}. Consider a fixed value of $u \in [1:K]$ and further consider the $u$-keys that are known to $i$ qualified receivers and $u-i$ eavesdropping receivers, where $i \in [1:u]$.

We focus on one specific set of $i$ qualified receivers, say receivers from the set $\mathcal{I}$ where $\mathcal{I} \subset [1:N], |\mathcal{I}| = i$. That is, we consider the keys $(s_{\mathcal{U}}: [1:N] \cap \mathcal{U} = \mathcal{I}, |\mathcal{U}| = u)$ and there are $\binom{K-N}{u-i}$ such $u$-keys. Further these $\binom{K-N}{u-i}$ keys are known to all qualified receivers from $\mathcal{I}$, and each eavesdropping receiver knows $\binom{K-N-1}{u-i-1}$ keys from these keys. Invoking generic linear codes similar to Example \ref{ex:sym}, we can securely send $(\binom{K-N}{u-i} - \binom{K-N-1}{u-i-1}) L^{[u]} = \binom{K-N-1}{u-i} L^{[u]}$ generic message symbols to receivers from $\mathcal{I}$ by transmitting $\binom{K-N-1}{u-i} L^{[u]}$ symbols.
\begin{eqnarray}
X^{[u],\mathcal{I}} = {\bf V}^w_{\mathcal{I}} W^{[u]} + \sum_{\mathcal{U}: [1:N] \cap \mathcal{U} = \mathcal{I}, |\mathcal{U}| = u} {\bf V}^s_{\mathcal{U}} s_{\mathcal{U}} \label{eq:syms}
\end{eqnarray}
where ${\bf V}^w_{\mathcal{I}}$ is a $\binom{K-N-1}{u-i} L^{[u]} \times \binom{N-1}{i-1} \binom{K-N-1}{u-i} L^{[u]}$ matrix over $\mathbb{F}_p$ and ${\bf V}^s_{\mathcal{U}}$ is a $\binom{K-N-1}{u-i} L^{[u]} \times L^{[u]}$ matrix over $\mathbb{F}_p$. Repeat the same coding procedure for all sets $\mathcal{I}$ such that $\mathcal{I} \subset [1:N]$ and $|\mathcal{I}| = i$. Consider the row stack of all the ${\bf V}^w_{\mathcal{I}}$ matrices appeared (denoted as ${\bf V}^{[u], w}$) and the column stack of all the ${\bf V}^s_{\mathcal{U}}$ matrices appeared (denoted as ${\bf V}^{[u], s}$). Set ${\bf V}^{[u], w}$ and ${\bf V}^{[u], s}$ as two Cauchy matrices from a sufficiently large field. The exact field size required and detailed analysis of security and correctness are similar to those in Theorem \ref{thm:unicast} and Theorem \ref{thm:multicast} and thus are not repeated here. Note that when we consider all sets of $i$ qualified receivers, overall there are $\binom{N}{i}$ choices and any particular qualified receiver is picked $\binom{N-1}{i-1}$ times, so the size of $W^{[u]}$ is set as  $\binom{N-1}{i-1} \binom{K-N-1}{u-i} L^{[u]}$. Security and correctness hold by the generic property of Cauchy matrices over large fields.

Counting all sets of $i$ qualified receivers and all $u$-keys, where $i \in [1:u], u \in [1:K]$, we calculate the overall performance as follows. 
\begin{eqnarray}
R &=&  \sum_{u=1}^K \sum_{i = 1}^u H(W^{[u]}) = \sum_{u=1}^K \sum_{i = 1}^u \binom{N-1}{i-1} \binom{K-N-1}{u-i} L^{[u]} \log_2 p \\
&=& \sum_{u=1}^K \sum_{j = 0}^{u-1} \binom{N-1}{j} \binom{K-N-1}{u-1-j} L^{[u]} \log_2 p = \sum_{u=1}^K \binom{K-2}{u-1} L^{[u]} \log_2 p
\end{eqnarray}
and 
\begin{eqnarray}
\beta(R) &=& \sum_{u=1}^K \sum_{i = 1}^u \binom{N}{i} \binom{K-N-1}{u-i} L^{[u]} \log_2 p \\
&=& \sum_{u=1}^K \left( \sum_{i = 0}^u \binom{N}{i} \binom{K-N-1}{u-i} - \binom{N}{0} \binom{K-N-1}{u} \right) L^{[u]} \log_2 p \\
&=& \sum_{u=1}^K \left( \binom{K-1}{u} - \binom{K-N-1}{u} \right) L^{[u]} \log_2 p 
\end{eqnarray}
where both rate and broadcast bandwidth match the converse bounds.
Note that we have used decompositions of schemes with independent $u$-keys (refer to Lemma \ref{lemma:decompose}) and for each $u$, we invoke the generic coding scheme (\ref{eq:syms}) where a specific finite field $\mathbb{F}_p$ is used. To ensure the overall scheme operates over the same field, we will use the maximum field size $p$ required for all component schemes and the Cauchy matrix based scheme in (\ref{eq:syms}) works for any field size that is larger than the minimum required. The proof of Theorem \ref{thm:sym} is thus complete.

\subsection{Proof of Theorem \ref{thm:52}: Rate Converse for The $N=2, K=5$ Instance}\label{sec:52}
The rate converse is split into two lemmas. Before presenting the lemmas, we first summarize the entropy identities from the problem description.
\begin{eqnarray}
\mbox{(Combinatorial Keys)} && H(a,b,c,d,e) = H(a) + H(b) + H(c) + H(d) + H(e) \label{eq:52h1}\\
\mbox{(Same Key Sizes)} && H(a) = H(b) = H(c) = H(d) = H(e) = L \label{eq:52h2}\\
\mbox{(Correctness)} && H(W|X, a, b, c) = H(W| X, b, d, e) = o(L) \label{eq:52corr}\\
\mbox{(Security)} && I(W; X, b) = I(W; X, c, d) = I(W; X, c, e) = o(L). \label{eq:52sec}
\end{eqnarray}
\begin{lemma}\label{lemma:521}
For the secure groupcast instance in Fig.~\ref{fig:52re}, we have
\begin{eqnarray}
H(d, e | W, X, b) \leq 2L - H(W) + o(L).
\end{eqnarray}
\end{lemma}

{\it Proof:}
\begin{eqnarray}
H(d, e | W, X, b) &=& H(d, e | X, b) - I(d, e; W | X, b) \\
&\overset{(\ref{eq:52h2})}{\leq}& 2L - H(W | X, b) + H(W| X, b, d, e) \\
&\overset{(\ref{eq:52sec})(\ref{eq:52corr})}{=}& 2L - H(W) + o(L).
\end{eqnarray}
\hfill \QED

\begin{lemma}\label{lemma:522}
For the secure groupcast instance in Fig.~\ref{fig:52re}, we have
\begin{eqnarray}
H(d, e | W, X, b) \geq 2H(W) - 3L + o(L).
\end{eqnarray}
\end{lemma}

{\it Proof:} Consider eavesdropping Receiver $4$ such that $X, c, d$ shall not reveal anything about $W$.
\begin{eqnarray}
H(W,X,c,d) &\overset{(\ref{eq:52sec})}{=}& H(W) + H(X, c, d)  + o(L) \\
&\geq& H(W) + H(X, b, c, d) - H(b) + o(L). \label{eq:521e}
\end{eqnarray}

Symmetrically, consider eavesdropping Receiver $5$ such that $X, c, e$ shall not reveal anything about $W$.
\begin{eqnarray}
H(W,X,c,e) &\overset{(\ref{eq:52sec})}{=}& H(W) + H(X, c, e)  + o(L) \\
&\geq& H(W) + H(X, b, c, e) - H(b) + o(L). \label{eq:522e}
\end{eqnarray}

Adding (\ref{eq:521e}) and (\ref{eq:522e}) and applying sub-modularity, we have
\begin{eqnarray}
&& H(W,X,c,d) + H(W,X,c,e) \notag\\
&\geq& 2H(W) -2H(b) + H(X, b, c, d, e) + H(X, b, c) + o(L)\\
&\overset{(\ref{eq:52corr})}{\geq}& 2 H(W) - 2H(b) + H(W, X, b, c, d, e) + H(X, a, b, c) - H(a) + o(L)\\
&\overset{(\ref{eq:52corr})}{\geq}& 2 H(W) - 2H(b) + H(W, X, c, d) + H(W, X, a, b, c) - H(a) + o(L).
\end{eqnarray}
Rearranging terms above and applying (\ref{eq:52h2}), we have
\begin{eqnarray}
2H(W) - 3L + o(L)&\leq& H(W, X, c, e) - H(W, X, a, b, c)  \\
&\leq& H(W, X, a, b, c, d, e) - H(W, X, a, b, c) \\
&=& H(d, e | W, X, a, b, c) \\
&\leq& H(d, e | W, X, b).
\end{eqnarray}
\hfill \QED

Finally, combining Lemma \ref{lemma:521} and Lemma \ref{lemma:522}, we have
\begin{eqnarray}
&& 2H(W) - 3L + o(L) \leq 2L - H(W) + o(L) \\
&\Rightarrow& R = \frac{H(W)}{L} \leq \frac{5}{3} + \frac{o(L)}{L} 
\end{eqnarray}
and letting $L \rightarrow \infty$ produces the desired bound $R \leq 5/3$.

\subsection{Proof of Theorem \ref{thm:region}: Multiple Messages}\label{sec:regionp}
Let us start with the converse proof, which is a generalization of that in Theorem \ref{thm:msg} and Theorem \ref{thm:bw}. Consider (\ref{eq:regionm1}) and (\ref{eq:regionm2}) follows from symmetry.
\begin{eqnarray}
(R_1 + R_{12}) L &=& H(W_1) + H(W_{12}) \\
&\overset{(\ref{eq:region1})}{=}& I(W_1, W_{12}; X, S_1, S_{12}) + o(L) \\
&\overset{(\ref{eq:region3})}{=}& I(W_1, W_{12}; S_1, S_{12} | X) + o(L) \\
&\leq& H(S_1, S_{12}) +o(L) = (H(s_1) + H(s_{12})) L + o(L).
\end{eqnarray}
Consider (\ref{eq:regionm3}) and (\ref{eq:regionm4}) follows from symmetry.
\begin{eqnarray}
R_1 L &=& H(W_1) \overset{(\ref{eq:region1})}{=} I(W_1; X, S_1, S_{12}) + o(L) \\
&\overset{(\ref{eq:region2})}{=}& I(W_1; S_1 | X, S_{12}) + o(L) \\
&\leq& H(S_1) +o(L) = H(s_1) L + o(L).
\end{eqnarray}
Consider (\ref{eq:regionb}).
\begin{eqnarray}
\beta(R_1, R_2, R_{12}) L &\geq& H(X) \geq H(X | S_1, S_2, S_{12}) \\
&\geq& I(X; W_1, W_2, W_{12} | S_1, S_2, S_{12}) \\
&\overset{(\ref{eq:region1})(\ref{eq:region2})}{=}& H(W_1, W_2, W_{12}|S_1, S_2, S_{12}) + o(L) \\
&\overset{(\ref{eq:regioni})}{=}& H(W_1) + H(W_2) + H(W_{12}) + o(L) \\
&=& (R_1 + R_2 + R_{12}) L + o(L), \\
\beta(R_1, R_2, R_{12}) &\geq& I(X; W_1, W_2, W_{12}, Z_1, Z_2) \\
&=& I(X; W_1, W_{12}, Z_1) + I(X; W_2, Z_2 | W_1, W_2, W_{12}, Z_1) \\
&\geq& I(X; W_1, W_{12} | Z_1) + I(X; Z_2 | W_1,W_2, W_{12}, Z_1) \\
&\overset{(\ref{eq:regioni})}{=}& I(X, Z_1; W_1, W_{12}) + I(X, W_2, W_{12}; Z_2 | W_1, Z_1)\\
&\overset{(\ref{eq:region1})}{=}& H(W_1, W_{12}) + I(X, W_2, W_{12}, Z_1; Z_2 | W_1) \notag\\
&&-I(Z_1; Z_2 | W_1) ~+ o(L)\\
&\overset{(\ref{eq:regioni})}{\geq}& H(W_1, W_{12}) + I(W_2, W_{12}; Z_2 | X, W_1) -I(Z_1; Z_2) + o(L)\\
&\overset{(\ref{eq:region2})}{=}& H(W_1, W_{12}) + H(W_2, W_{12} | X, W_1) -I(Z_1; Z_2) + o(L)\\
&\overset{(\ref{eq:region3})}{=}& H(W_1, W_{12}) + H(W_2, W_{12} | W_1) -I(Z_1; Z_2) + o(L)\\
&\overset{(\ref{eq:regioni})}{=}& (R_1 + R_{12} + R_2 + R_{12} - H(s_{12}))L + o(L).
\end{eqnarray}
Next, we consider the achievability. Consider any rational rate tuple $(R_1, R_2, R_{12}) \in \mathcal{C}$, i.e., $(R_1, R_2, R_{12})$ satisfies (\ref{eq:regionm1}) - (\ref{eq:regionm4}). Without loss of generality, assume $R_1, R_2, R_{12}$ are integers (for rationals, we may consider blocks over the least common multiple of the denominators so that the number of bits becomes integers). We operate over the binary field and consider $L=1$ block. The transmit signal is designed as follows. Denote by $W^{[a_1:a_2]}$ the $a_1$-th to $a_2$-th bits in the vector $W$. We have two cases.
\begin{enumerate}
\item[] Case 1. $R_{12} \leq H(s_{12})$.
\begin{eqnarray}
X = (
W_1 + s_1^{[1:R_1]};~
W_2 + s_2^{[1:R_2]};~
W_{12} + s_{12}^{[1: R_{12}]}
).
\end{eqnarray}
The broadcast bandwidth achieved is $\beta(R_1, R_2, R_{12}) = R_1 + R_2 + R_{12}$.
\item[] Case 2. $R_{12} > H(s_{12})$.
\begin{eqnarray}
X = \left(
\begin{array}{c}
W_1 + s_1^{[1:R_1]};~
W_2 + s_2^{[1:R_2]};~
W_{12}^{[1: H(s_{12})]} + s_{12} \\
W_{12}^{[H(s_{12})+1:R_{12}]} + s_1^{[R_1+1 : R_1+R_{12}- H(s_{12})]} \\
W_{12}^{[H(s_{12})+1:R_{12}]} + s_2^{[R_2+1 : R_2+R_{12}- H(s_{12})]} \\
\end{array}
\right).
\end{eqnarray}
Note that as $(R_1, R_2, R_{12})$ satisfies $(\ref{eq:regionm1}) - (\ref{eq:regionm4})$, the key bits in the above scheme exist. The broadcast bandwidth achieved is $\beta(R_1, R_2, R_{12}) = R_1 + R_2 + 2R_{12} - H(s_{12})$.
\end{enumerate}
Thus any rational rate tuple in the capacity region is achievable and as rational tuples are dense over the reals, the proof of Theorem \ref{thm:region} is complete.

\subsection{Proof of Theorem \ref{thm:dmc}: Achievability under Discrete Memoryless Keys}\label{sec:dmcp}
Before presenting the achievability proof for Theorem \ref{thm:dmc}, we cite a lemma on privacy amplification\footnote{Lemma \ref{lemma:pa} on secret key extraction suffices for our purposes over long key block lengths. Stronger non-asymptotic results and more efficient constructions of the random mappings are available in the literature (see e.g., \cite{Bennett_Brassard_Crepeau_Maurer, Hayashi_MinEntropy}).}, which encapsulates most technicalities of the achievability proof.
\begin{lemma}\label{lemma:pa}
{\normalfont(Lemma 5.18 in \cite{Narayan_Tyagi})}
Consider random variables $Z_c, Z_e, X_e$ (with finite cardinality) such that $Z_c, Z_e$ are $L$ length extensions of $z_c, z_e$ and $L_{X_e}$ denotes the number of bits in $X_e$. Then there exists a random mapping (independent of $Z_e, X_e$) from $Z_c$ to a uniform random variable $Z$ with $L_Z$ bits such that
\begin{eqnarray}
&& L_Z = H(z_c | z_e)L - L_{X_e} + o(L), \label{eq:pal}\\
&& I(Z; Z_e, X_e) = o(L). \label{eq:pasec}
\end{eqnarray}
\end{lemma}

In Lemma \ref{lemma:pa}, we may interpret $Z$ as the secret key to be generated from a known variable $Z_c$ such that $Z$ is almost independent of an eavesdropped variable $Z_e$ (that has certain joint distribution with $Z_c$) and a prior knowledge variable $X_e$ (that is arbitrarily correlated with $Z_c, Z_e$). The secret key size turns out to be given by the conditional entropy value minus the leaked prior knowledge.

Consider secure unicast first, whose achievability proof follows immediately from Lemma \ref{lemma:pa}. Set $Z_c = Z_1$, i.e., the key for qualified Receiver $1$, $Z_e$ as the key for eavesdropping Receiver $e \in [2:K]$, and $X_e = ()$. From Lemma \ref{lemma:pa}, we know that Receiver $1$ can generate a key $Z$ that is almost independent from any eavesdropping receiver. Note that the random mapping used in Lemma \ref{lemma:pa} does not depend on the eavesdropped variable $Z_e$ so that the secret key $Z$ generated is simultaneously independent of any eavesdropped variable as long as we pick the key length to be $L_Z = \min_{e \in [2:K]} H(z_1|z_e) L + o(L)$.
We use the key to send the desired message through one-time pad, i.e., $X = W + Z$ where the length of $W$ is the same as the length of $Z$. Correctness and security are easy to verify (as $Z$ is almost independent of $Z_e$, see (\ref{eq:pasec})). The rate and broadcast bandwidth achieved are $R = \beta(R) = \min_{e\in[2:K]} H(z_1|z_e)$ as $L \rightarrow \infty$.

Then consider secure multicast. In Lemma \ref{lemma:pa}, we set $Z_c = (Z_1, Z_2, \cdots, Z_{K-1})$, $Z_e = Z_K$, and $X_e$ as the random bin index of $(Z_1, Z_2, \cdots, Z_{K-1})$ of length $\max_{q\in[1:K-1]}H(z_1,\cdots,z_{K-1} | z_q, z_K)L + o(L)$. The key $Z$ is generated from $Z_c$ and has length as specified in Lemma \ref{lemma:pa}. The transmit signal sent by the transmitter is $X = (Z_K, X_e, W+Z)$. From $Z_K, X_e$, every qualified Receiver $q, q \in [1:K-1]$ can recover $Z_c$ (by Slepian Wolf coding as the overall information seen by Receiver $q$, i.e., $Z_K, X_e, Z_q$ has entropy whose value is at least the joint entropy, $H(Z_1, \cdots, Z_K)$) and then generate $Z$ with the same random mapping used by the transmitter. After extracting the common key $Z$, $W$ can be decoded with vanishing error by every qualified receiver. Security is guaranteed by Lemma \ref{lemma:pa} as $Z$ is almost independent of the information available to the eavesdropping Receiver $K$, i.e., $Z_K, X_e$. The rate achieved is $R = H(z_1, \cdots, z_{K-1} | z_K) - \max_{q\in[1:K-1]}H(z_1,\cdots,z_{K-1} | z_q, z_K) = \min_{q\in[1:K-1]} H(z_q|z_K)$ as $L \rightarrow \infty$.

\section{Conclusion}
We introduce the problem of secure groupcast, where a transmitter wishes to securely communicate with a group of selected receivers while ensuring the other illegitimate receivers are fully ignorant of the desired communication, through noiseless broadcasting and correlated keys.

The communication efficiency of secure groupcast is measured by the message rate (number of message bits securely groupcast) and the broadcast bandwidth resource used (number of bits in the transmit signal). The main emphasis is placed on the most elementary setting of combinatorial keys and one common message, and limited extensions are also explored. Complete answers are obtained for certain preliminary cases, e.g., one legitimate receiver or one eavesdropping receiver, symmetric cases, while other cases remain unsolved. Interesting insights emerge out of this study, e.g., the necessity of decomposition and both structured and generic coding, the quest for tighter general converse bounds, and the potential of alignment view of the correlated key, message and transmit signal spaces. We find secure groupcast to be an interesting and challenging information theoretic security primitive with many open questions, and this work is a first step towards understanding coding opportunities for group communications under multiple correlated eavesdroppers.

\let\url\nolinkurl
\bibliographystyle{IEEEtran}
\bibliography{Thesis}

\begin{thebibliography}{10}
\providecommand{\url}[1]{#1}
\csname url@samestyle\endcsname
\providecommand{\newblock}{\relax}
\providecommand{\bibinfo}[2]{#2}
\providecommand{\BIBentrySTDinterwordspacing}{\spaceskip=0pt\relax}
\providecommand{\BIBentryALTinterwordstretchfactor}{4}
\providecommand{\BIBentryALTinterwordspacing}{\spaceskip=\fontdimen2\font plus
\BIBentryALTinterwordstretchfactor\fontdimen3\font minus
  \fontdimen4\font\relax}
\providecommand{\BIBforeignlanguage}[2]{{%
\expandafter\ifx\csname l@#1\endcsname\relax
\typeout{** WARNING: IEEEtran.bst: No hyphenation pattern has been}%
\typeout{** loaded for the language `#1'. Using the pattern for}%
\typeout{** the default language instead.}%
\else
\language=\csname l@#1\endcsname
\fi
#2}}
\providecommand{\BIBdecl}{\relax}
\BIBdecl

\bibitem{Shannon1949}
C.~E. Shannon, ``{Communication Theory of Secrecy Systems},'' \emph{Bell system
  technical journal}, vol.~28, no.~4, pp. 656--715, 1949.

\bibitem{Li_Sun_CDS}
Z.~Li and H.~Sun, ``{Conditional Disclosure of Secrets: A Noise and Signal
  Alignment Approach},'' \emph{arXiv preprint arXiv:2002.05691}, 2020.

\bibitem{Zhou_Sun_Fu}
Y.~{Zhou}, H.~{Sun}, and S.~{Fu}, ``{On the Randomness Cost of Linear Secure
  Computation},'' in \emph{2019 53rd Annual Conference on Information Sciences
  and Systems (CISS)}, March 2019, pp. 1--6.

\bibitem{Sun_Jafar_PIR}
H.~Sun and S.~A. Jafar, ``{The Capacity of Private Information Retrieval},''
  \emph{IEEE Transactions on Information Theory}, vol.~63, no.~7, pp.
  4075--4088, 2017.

\bibitem{Sudan_Tyagi_Watanabe}
M.~Sudan, H.~Tyagi, and S.~Watanabe, ``{Communication for Generating
  Correlation: A Unifying Survey},'' \emph{IEEE Transactions on Information
  Theory}, vol.~66, no.~1, pp. 5--37, 2019.

\bibitem{Csiszar_Narayan}
I.~Csiszar and P.~Narayan, ``{Secrecy Capacities for Multiple Terminals},''
  \emph{IEEE Transactions on Information Theory}, vol.~50, no.~12, pp.
  3047--3061, 2004.

\bibitem{Slepian_Wolf}
D.~Slepian and J.~Wolf, ``{Noiseless Coding of Correlated Information
  Sources},'' \emph{IEEE Transactions on information Theory}, vol.~19, no.~4,
  pp. 471--480, 1973.

\bibitem{Maurer_Key}
U.~M. Maurer, ``{Secret Key Agreement by Public Discussion from Common
  Information},'' \emph{IEEE Transactions on Information Theory}, vol.~39,
  no.~3, pp. 733--742, 1993.

\bibitem{Ahlswede_Csiszar_CR}
R.~Ahlswede and I.~Csiszar, ``{Common Randomness in Information Theory and
  Cryptography - Part I: Secret Sharing},'' \emph{IEEE Transactions on
  Information Theory}, vol.~39, no.~4, pp. 1121--1132, 1993.

\bibitem{Gohari_Anantharam}
A.~A. Gohari and V.~Anantharam, ``{Information-Theoretic Key Agreement of
  Multiple Terminals - Part I},'' \emph{IEEE Transactions on Information
  Theory}, vol.~56, no.~8, pp. 3973--3996, 2010.

\bibitem{Chan_Zheng}
C.~Chan and L.~Zheng, ``{Mutual Dependence for Secret Key Agreement},'' in
  \emph{2010 44th Annual Conference on Information Sciences and Systems
  (CISS)}.\hskip 1em plus 0.5em minus 0.4em\relax IEEE, 2010, pp. 1--6.

\bibitem{Grokop_Tse}
L.~Grokop and D.~Tse, ``{Fundamental Constraints on Multicast Capacity
  Regions},'' in \emph{2007 45th Annual Allerton Conference on Communication,
  Control, and Computing (Allerton)}.

\bibitem{Tian_Latent}
C.~Tian, ``{Latent Capacity Region: A Case Study on Symmetric Broadcast with
  Common Messages},'' \emph{IEEE Transactions on Information Theory}, vol.~57,
  no.~6, pp. 3273--3285, 2011.

\bibitem{Salimi_Liu_Cui}
A.~Salimi, T.~Liu, and S.~Cui, ``{Polyhedral Description of the Symmetrical
  Latency Capacity Region of Broadcast Channels},'' in \emph{2014 IEEE
  International Symposium on Information Theory}.\hskip 1em plus 0.5em minus
  0.4em\relax IEEE, 2014, pp. 2122--2126.

\bibitem{Wyner_Wiretap}
A.~D. Wyner, ``{The Wire-Tap Channel},'' \emph{Bell system technical journal},
  vol.~54, no.~8, pp. 1355--1387, 1975.

\bibitem{Csiszar_Korner}
I.~Csiszar and J.~Korner, ``{Broadcast Channels with Confidential Messages},''
  \emph{IEEE Transactions on Information Theory}, vol.~24, no.~3, pp. 339--348,
  1978.

\bibitem{Liu_Maric_Spasojevic_Yates}
R.~Liu, I.~Maric, P.~Spasojevic, and R.~D. Yates, ``{Discrete Memoryless
  Interference and Broadcast Channels with Confidential Messages: Secrecy Rate
  Regions},'' \emph{IEEE Transactions on Information Theory}, vol.~54, no.~6,
  pp. 2493--2507, 2008.

\bibitem{Khisti_Tchamkerten_Wornell}
A.~Khisti, A.~Tchamkerten, and G.~W. Wornell, ``{Secure Broadcasting over
  Fading Channels},'' \emph{IEEE Transactions on Information Theory}, vol.~54,
  no.~6, pp. 2453--2469, 2008.

\bibitem{Ekrem_Ulukus}
E.~Ekrem and S.~Ulukus, ``{Secure Broadcasting Using Multiple Antennas},''
  \emph{Journal of Communications and Networks}, vol.~12, no.~5, pp. 411--432,
  2010.

\bibitem{Schaefer_Khisti_Poor}
R.~F. Schaefer, A.~Khisti, and H.~V. Poor, ``{Secure Broadcasting Using
  Independent Secret Keys},'' \emph{IEEE Transactions on Communications},
  vol.~66, no.~2, pp. 644--661, 2017.

\bibitem{Yossef_Birk_Jayram_Kol_Trans}
{Z. Bar-Yossef and Y. Birk and T. S. Jayram and T. Kol}, ``{Index Coding With
  Side Information},'' \emph{IEEE Trans. on Information Theory}, vol.~57,
  no.~3, pp. 1479 -- 1494, March 2011.

\bibitem{Dau_Skachek_Chee_Security}
S.~H. Dau, V.~Skachek, and Y.~M. Chee, ``{On the Security of Index Coding with
  Side Information},'' \emph{IEEE Transactions on Information Theory}, vol.~58,
  no.~6, pp. 3975--3988, 2012.

\bibitem{Mojahedian_Aref_Gohari}
M.~M. Mojahedian, M.~R. Aref, and A.~Gohari, ``{Perfectly Secure Index
  Coding},'' \emph{IEEE Transactions on Information Theory}, vol.~63, no.~11,
  pp. 7382--7395, 2017.

\bibitem{Narayanan_Prabhakaran_Ravietal}
V.~Narayanan, V.~M. Prabhakaran, J.~Ravi, V.~K. Mishra, B.~K. Dey, and
  N.~Karamchandani, ``{Private Index Coding},'' in \emph{2018 IEEE
  International Symposium on Information Theory (ISIT)}.\hskip 1em plus 0.5em
  minus 0.4em\relax IEEE, 2018, pp. 596--600.

\bibitem{Bennett_Brassard_Crepeau_Maurer}
C.~H. Bennett, G.~Brassard, C.~Cr{\'e}peau, and U.~M. Maurer, ``{Generalized
  Privacy Amplification},'' \emph{IEEE Transactions on Information Theory},
  vol.~41, no.~6, pp. 1915--1923, 1995.

\bibitem{Hayashi_MinEntropy}
M.~Hayashi, ``{Security Analysis of $\epsilon$-Almost Dual $\mbox{Universal}_2$
  Hash Functions: Smoothing of Min Entropy Versus Smoothing of R{\'e}nyi
  Entropy of Order 2},'' \emph{IEEE Transactions on Information Theory},
  vol.~62, no.~6, pp. 3451--3476, 2016.

\bibitem{Narayan_Tyagi}
P.~Narayan and H.~Tyagi, ``{Multiterminal Secrecy by Public Discussion},''
  \emph{Foundations and Trends in Communications and Information Theory},
  vol.~13, no. 2-3, pp. 129--275, 2016.

\end{thebibliography}
\end{document}